\documentclass[letterpaper,twocolumn,10pt]{article}
\usepackage{usenix-2020-09}

\usepackage{tikz}
\usepackage{amsmath}

\usepackage{algorithmic}
\usepackage[ruled,linesnumbered]{algorithm2e}
\usepackage{url}
\usepackage{booktabs}
\usepackage{fancyhdr}
\usepackage{graphicx}
\usepackage{float}
\usepackage{subfig}
\usepackage{authblk}
\SetKwProg{Fn}{Function}{}{end}\SetKwFunction{FRecurs}{FnRecursive}%
\SetKwComment{Comment}{// }{ }
\definecolor{commentColor}{RGB}{106,153,81}

\SetCommentSty{mycommfont}

\newcommand*\circled[1]{\tikz[baseline=(char.base)]{
\node[shape=circle,fill,inner sep=0.2pt] (char) {\textcolor{white}{#1}};}}

\begin{document}

\date{}


\title{\Large \bf FusionANNS: An Efficient CPU/GPU Cooperative Processing Architecture for Billion-scale Approximate Nearest Neighbor Search}

\author{
{\rm Bing Tian$^{\dag}$, Haikun Liu$^{\dag}$, Yuhang Tang$^{\dag}$, Shihai Xiao$^{\ddag}$, Zhuohui Duan$^{\dag}$, Xiaofei Liao$^{\dag}$, Xuecang Zhang$^{\ddag}$, Junhua Zhu$^{\ddag}$, Yu Zhang$^{\dag}$}\\ $^{\dag}$National Engineering Research Center for Big Data Technology and System,\\
Service Computing Technology and System Lab/Cluster and Grid Computing Lab,\\
School of Computer Science and Technology, Huazhong University of Science and Technology, China\\
$^{\ddag}$Huawei Technologies Co., Ltd\\}

\maketitle

\begin{abstract}
\textit{Approximate nearest neighbor search} (ANNS) has emerged as a crucial component of database and AI infrastructure. Ever-increasing vector datasets pose significant challenges in terms of performance, cost, and accuracy for ANNS services. None of  modern ANNS systems can address these issues  simultaneously. 

We present FusionANNS, a high-throughput, low-latency, cost-efficient, and high-accuracy ANNS system for billion-scale datasets using SSDs and only one entry-level GPU. The key idea of FusionANNS lies in CPU/GPU collaborative filtering and re-ranking mechanisms, which significantly reduce I/O operations across CPUs, GPU, and SSDs to break through the I/O performance bottleneck. Specifically, we propose three novel designs: (\textit{1}) \textit{multi-tiered indexing} to avoid  data swapping between CPUs and GPU, (\textit{2}) \textit{heuristic re-ranking} to eliminate unnecessary I/Os and computations while guaranteeing high accuracy, and (\textit{3}) \textit{redundant-aware I/O deduplication} to further improve I/O efficiency. We implement FusionANNS and compare it with the state-of-the-art SSD-based ANNS system--SPANN and GPU-accelerated in-memory ANNS system--RUMMY. Experimental results show that FusionANNS achieves 1) 9.4-13.1$\times$ higher \textit{query per second} (QPS) and 5.7-8.8$\times$ higher cost efficiency compared with SPANN; 2) and 2-4.9$\times$ higher QPS and 2.3-6.8$\times$ higher cost efficiency compared with RUMMY, while guaranteeing low latency and high accuracy.



\end{abstract}

\section{Introduction}
\begin{figure}[t]
  \centering
  \includegraphics[clip,width=\linewidth]{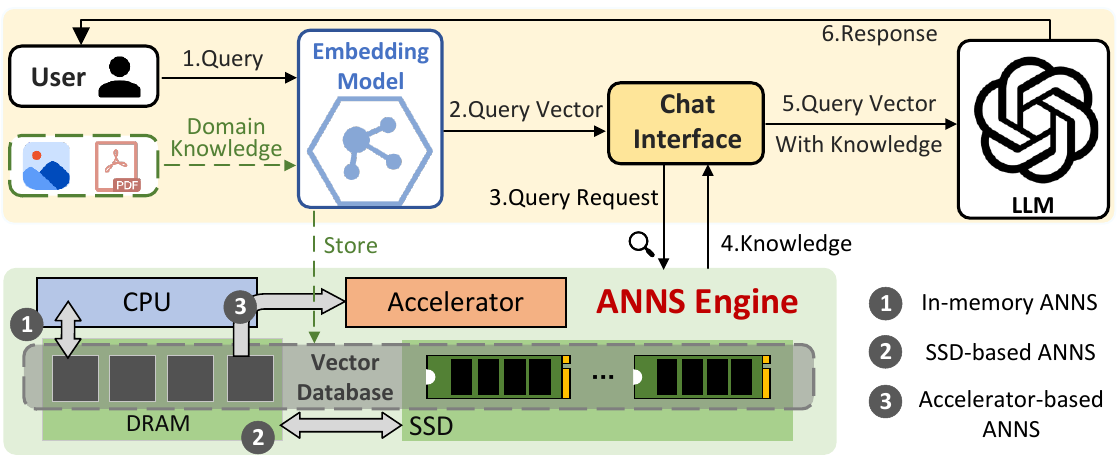}
  \vspace{-4.5ex}
  \caption{The framework of retrieval augmented generation}
  \vspace{-3ex}
  \label{ChatGPT}
\end{figure}


\textit{Approximate nearest neighbor search} (ANNS) in high-dimensional spaces refers to find top-$k$ vectors most similar to a given query vector.  
It has a wide range of applications in many fields, including data mining~\cite{datamining}, search engines~\cite{BingKDD22}, and AI-driven recommendation systems~\cite{AlibabaKDD18, facebookKDD20}. Specifically, fueled by the recent prosperity of \textit{Large Language Models} (LLMs)~\cite{LLM1,LLM2,LLM3,LLM4},  ANNS systems have become a crucial component of modern AI infrastructure. Figure~\ref{ChatGPT} shows a typical framework of \textit{Retrieval Augmented Generation} (RAG). The domain-specific knowledge is first embedded as high-dimensional vectors and stored in a vector database. When a chatbot receives a query, it uses the ANNS engine to retrieve the most relevant knowledge from the vector database, allowing the LLM to use that knowledge as additional context for more accurate inference.

ANNS is a typical \textbf{memory-hungry} and \textbf{compute-intensive} application. Most ANNS systems~\cite{RUMMY,Bang,DF-GAS,ParlayANN,iQAN} exploit \textit{inverted file} (IVF)~\cite{SPFresh,SPANN,IVF_1} or graph-based~\cite{ParlayANN,iQAN,HNSW} indices to facilitate ANNS. For billion-scale datasets, these indices  usually require a large amount of memory resource. For example, state-of-the-art IVF-based RUMMY~\cite{RUMMY} and graph-based Bang~\cite{Bang} require terabyte-scale memory space to accommodate billion-scale vectors and their indices. The substantial memory demand significantly increases the \textit{total cost of ownership} (TCO), impeding ANNS scaling to extremely large datasets (e.g., hundreds of billions of vectors).  Despite the huge memory requirement, ANNS is
also computationally intensive because it requires massive distance calculations among vectors, especially for large-scale datasets in high-dimensional spaces. With a rapid growth of the vector database, ANNS has emerged as a new performance bottleneck in RAG scenarios~\cite{RUMMY}, potentially accounting for about 50\% of the total latency for an LLM query~\cite{RAG}.

To reduce the cost of memory required by ANNS, there are mainly two kinds of approaches, i.e., \textit{Hierarchical Indexing} (HI)~\cite{SPANN,SmartANNS} and \textit{Product Quantization} (PQ)~\cite{Faiss_GPU,JUNO}. \textbf{First, the hierarchical indexing approach reduces memory consumption by  storing  
indices~\cite{SPANN,DiskANN} on SSDs}. 
Typically, Microsoft's commercial ANNS system--SPANN~\cite{SPANN, SPFresh} stores all IVF-based indices (i.e., posting lists) on SSDs, and maintains the centroids of these posting lists in memory using a navigation graph. 
Although SPANN achieves low latency, 
we find that its throughput for concurrent queries is quite limited, peaking at only four CPU threads on a high-end SSD (Section~\ref{SPANN limited}). The limited scalability hampers its practicality for AI applications requiring high throughput.
\textbf{Second, PQ is another effective way for memory cost saving}. This vector compression technology can significantly  reduce the memory footprint of high-dimensional vectors by up to 95\%, and can also accelerate the ANNS speed by several times~\cite{Faiss_GPU}. However, since PQ is a lossy-compression scheme, a higher compression rate often implies a lower query accuracy. It is usually unacceptable for some scenarios that require high accuracy (e.g. recall$\geq90\%$)~\cite{RAGSurvey}. 

To address the computing challenge, GPUs have been increasingly leveraged to accelerate extensive distance calculations involved in ANNS. Recent GPU-based ANNS solutions~\cite{RUMMY,Bang,GPU_Graph1, GPU_Graph2,GPU_Graph3,Faiss_GPU} have demonstrated high efficiency for handling small datasets that fit within the GPU's \textit{high bandwidth memory} (HBM). However, for billion-scale datasets, the GPU-based approach may suffer from significant performance degradation. 
Our experiments show that the performance of ANNS even declines by 10\% when SPANN directly adopts  GPUs for distance calculations (Section~\ref{SPANN variant}). The root cause is that the limited capacity of HBM causes extensive data movement across GPU's HBM, host memory, and SSDs. 




Although the above approaches can address some of the performance/cost/accuracy issues to some extent, none of them can offer  high throughput, low latency, cost efficiency, and high accuracy simultaneously for billion-scale ANNS services. Intuitively, one can adopt  hierarchical indexing, product quantization, and GPU acceleration techniques together to achieve an optimal ANNS solution. However, we find that the combination of these techniques causes even worse performance than SPANN which exploits hierarchical indexing solely (Section~\ref{SPANN variant}). Overall, there remains several challenges to collaborate hierarchical indexing with product quantization in a GPU-accelerated ANNS system.


\textbf{Challenge 1:}  To improve query accuracy and efficiency, most ANNS systems~\cite{SPFresh,SPANN} exploit a replication strategy to build high-quality IVF indices, where boundary vectors are replicated into adjacent posting lists. This can significantly expand the size of indices by $8\times$ larger than that of raw vectors~\cite{SPFresh,SPANN}. Even these indices are  compressed with PQ, the GPU's HBM still cannot accommodate all compressed indices, resulting in extensive data swapping between GPU and CPUs. \textbf{Challenge 2:} Since PQ incurs non-trivial accuracy loss, it is often associated with a vector re-ranking process to improve the query accuracy. However, since the accuracy loss varies significantly among different compressed vectors, it is challenging to determine the minimum number of vectors that requires re-ranking for each query  under a given accuracy constrain. 
\textbf{Challenge 3:} Since a raw vector (128$\sim$384 bytes) is much smaller than the minimum read granularity (4 KB) of modern NVMe SSDs, each request for raw vectors often causes significant read amplification, resulting in low I/O efficiency during re-ranking.

In this paper, we present FusionANNS, a ``CPU + GPU'' cooperative processing
architecture for billion-scale ANNS. 
FusionANNS achieves high throughput, low latency, cost efficiency and high accuracy simultaneously using only one entry-level GPU. The key idea of FusionANNS is to minimize data swapping across  GPU, CPUs, and SSDs via CPU/GPU collaborative filtering and re-ranking. Specifically, we propose three novel designs to tackle the above challenges. 

First, we propose a novel \textit{multi-tiered index structure} to enable CPU/GPU collaborative filtering. FusionANNS stores (\textit{i}) raw vectors on SSDs, and (\textit{ii}) compressed vectors using PQ in the GPU's HBM, while  maintaining (\textit{iii}) only vector-IDs of each posting list and a navigation graph in host memory. Since the HBM only stores highly-compressed PQ-vectors rather than compressed posting lists, it can accommodate all compressed vectors in billion-scale datasets. Upon a query, the host CPU first traverses the in-memory navigation graph to find the top-$m$ nearest posting lists, and then only transmit their vector-IDs (excluding the vectors' content) to GPU for distance calculations. In this way, FusionANNS can significantly reduce data transmission between CPUs and GPU.

Second, we propose  \textit{heuristic re-ranking} to improve the query accuracy while avoiding unnecessary I/O operations and distance calculations. We split the re-ranking process into multiple mini-batches and execute them sequentially. Once a mini-batch is finished, we exploit a lightweight feedback control model to check whether subsequent mini-batches are beneficial for improving the query accuracy, and  terminate the re-ranking process immediately if successive mini-batches have little contribute to the query accuracy.   

Third, we propose  \textit{redundancy-aware I/O deduplication} to further improve the I/O efficiency during re-ranking. We store vectors with high similarity compactly to improve the spatial locality on SSDs. This optimized storage layout enables two I/O deduplication mechanisms: 1) merging multiple I/Os mapped to the same page of SSDs within a mini-batch to mitigate read amplification, 2) fully exploiting the DRAM buffer to eliminate redundant I/Os in  subsequent mini-batches. 

Overall, we make the following contributions:

\begin{itemize}
\vspace{-1.5ex}
    \item We design FusionANNS, the first GPU-accelerated SSD-based ANNS system that achieves  high throughput, low latency, cost efficiency and high accuracy simultaneously for billion-scale datasets.
    \vspace{-1.5ex}
    \item For \textbf{Challenge 1}, we propose a novel \textit{multi-tiered index} that enables  GPU/CPU collaborative filtering to  significantly reduce data transmission between CPUs and GPU.
    \vspace{-1.5ex}
    \item For \textbf{Challenge 2}, we propose  \textit{heuristic re-ranking} to eliminate unnecessary I/Os and computations during re-ranking.
    \vspace{-1.5ex}
    \item For \textbf{Challenge 3}, we propose \textit{redundancy-aware I/O deduplication} based on the optimized storage layout to further enhance I/O efficiency.
    \vspace{-1.5ex}
    \item We evaluate FusionANNS using a general purpose server equipped with an entry-level GPU. Experimental results show that FusionANNS improves QPS by up to 13.1\(\times\) and 4.9\(\times\), and enhance cost efficiency by up to 8.8\(\times\) and 6.8\(\times\), compared with the state-of-the-art SSD-based system--SPANN and GPU-accelerated in-memory system--RUMMY, respectively, while  guaranteeing low latency and high accuracy. 
\end{itemize}

\section{Background and Motivation}
In this section, we first introduce two kinds of ANNS indexing techniques and product quantization (PQ) for vectors, and then 
present our main idea and analyze its key challenges, which motivate the design of FusionANNS.

\subsection{Indexing Techniques for ANNS}
\label{SPANN limited}
Most ANNS algorithms exploit a distance metric such as Euclidean distance to find the top-k nearest neighbors for a given query vector. For high-dimensional and large-scale datasets, it is computationally costly due to the curse of dimensionality~\cite{dimensionality}. To address this issue, most ANNS algorithms~\cite{hash1,hash2,Tree,HNSW,IVF_1,IVF_2,IVF-PQ} exploit indexing techniques to prune data regions that are unlikely to contain the nearest neighbors. These indices can significantly improve the query performance by shrinking the search space, but significantly increases memory consumption, especially for large datasets. Among various indexing techniques, IVF~\cite{IVF_1,IVF_2,IVF-PQ,SPANN} and graph-based~\cite{HNSW,NSG} indices are widely used due to their high efficiency. 

The \textbf{graph-based index} often organizes vectors in a proximity graph structure, in which vertices and edges represent vectors and distances between two vertices, respectively.
Upon a query, the ANNS engine traverses the graph from a given vertex to find the top-$k$ nearest neighbors. DiskANN~\cite{DiskANN} is a typical graph-based ANNS solution. It uses SSDs to store graph indices of billion-scale datasets while keeping some frequently-accessed vertices in main memory. Although DiskANN is a memory-efficient ANNS solution, it experiences high latency for queries due to rather long iteration paths for large-scale datasets.

\begin{figure}[t]
  \centering
  \includegraphics[clip,width=0.85\linewidth]{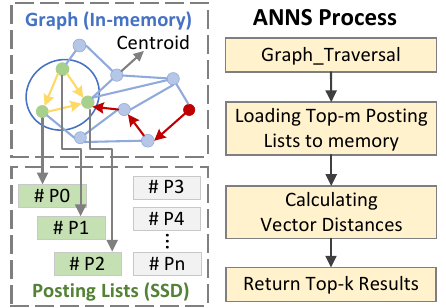}
  \vspace{-1ex}
  \caption{The hierarchical indexing technique in SPANN}
  \vspace{-3ex}
  \label{SPANN_algorithm}
\end{figure}


      \begin{figure} [t] 
      \centering  
      \subfloat[Throughput]{  
        \includegraphics[clip, width=0.48\linewidth]{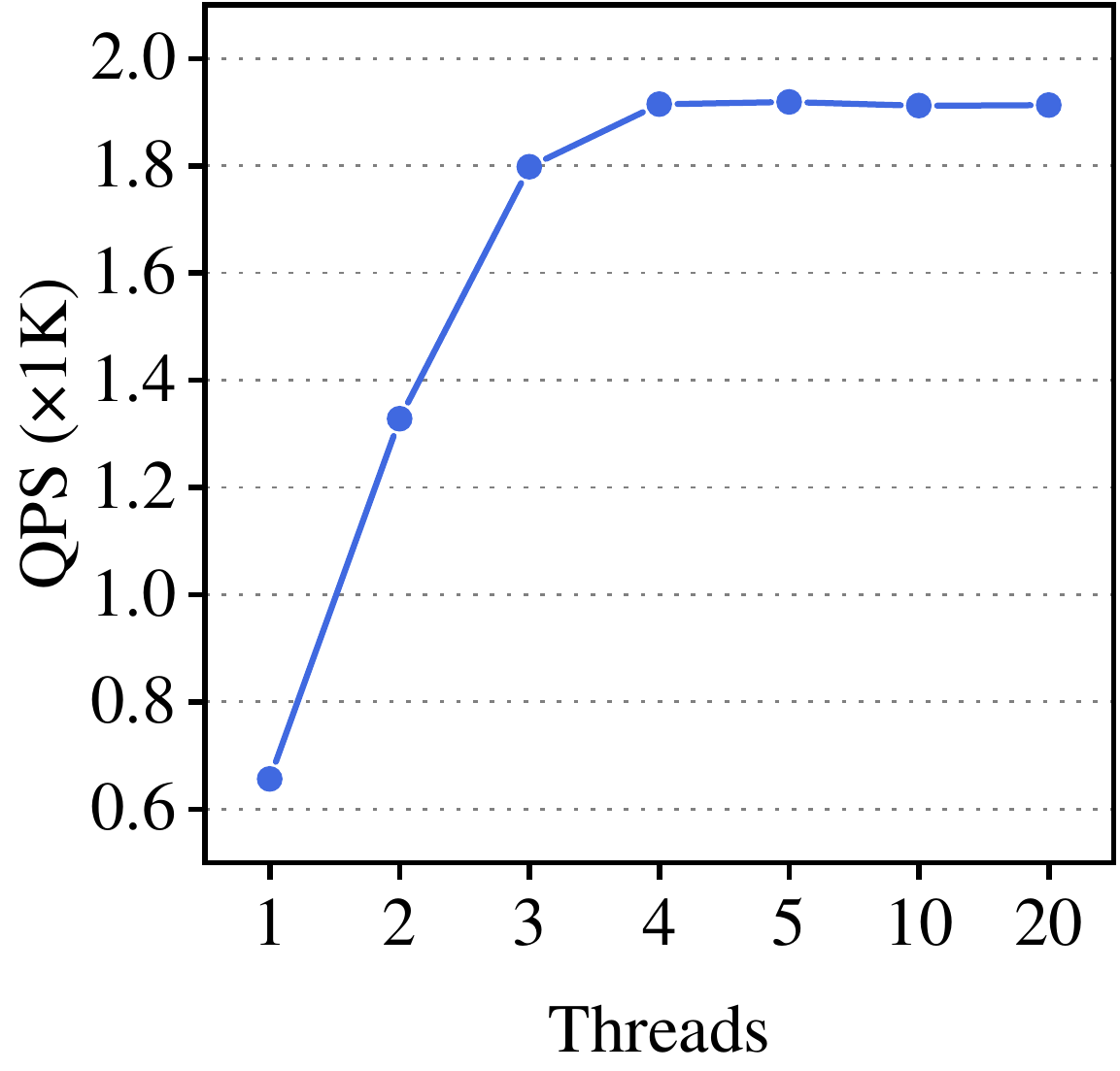}  
        \label{throughput}  
      } 
      \subfloat[Latency]{  
        \includegraphics[clip, width=0.48\linewidth]{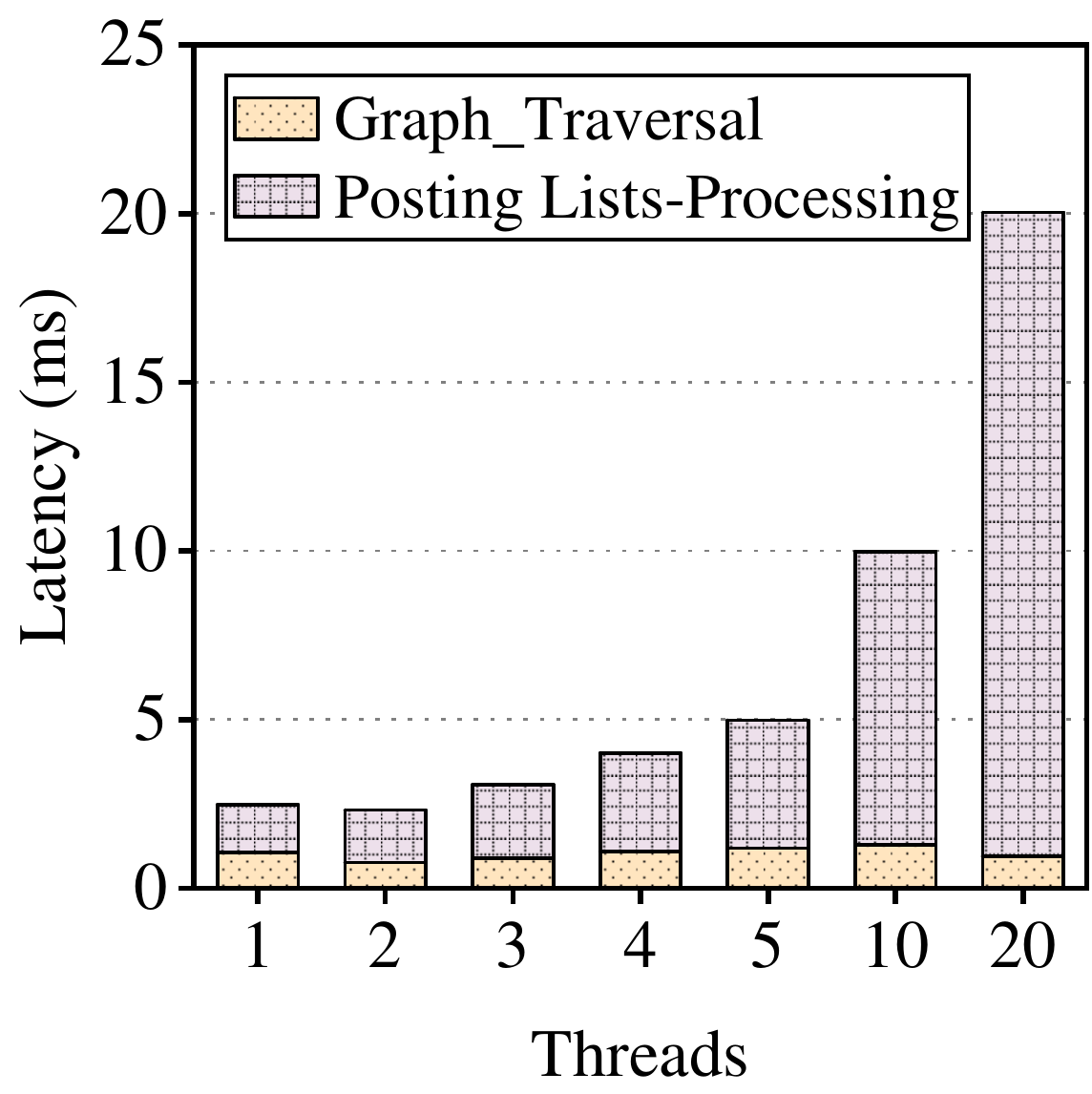}  
        \label{latency}  
      }  
      \vspace{-2ex}  
      \caption{ The throughput and  latency of SPANN
      } 
      \label{spann performance}
      \vspace{-3ex} 
      \end{figure}

The \textbf{IVF index} is a popular indexing technique for large-scale datasets stored on SSDs. To create the IVF index, a dataset is often partitioned into many posting lists using a clustering algorithm~\cite{cluster}, and each posting list is represented by its centroid. Recent studies~\cite{RUMMY,IndexStudy} have demonstrated that the IVF index~\cite{SPANN} is more efficient than the state-of-the-art graph-based index~\cite{DiskANN} for billion-scale datasets. 
SPANN~\cite{SPANN} is a state-of-the-art billion-scale ANNS system using the IVF index. Unlike  conventional IVF indices, SPANN builds an advanced IVF index by replicating boundary vectors of clusters into adjacent posting lists. This replication mechanism significantly expands the size of vector indices by $8\times$, but improves the query accuracy and efficiency. As shown in Figure~\ref{SPANN_algorithm}, SPANN stores all posting lists on SSDs while maintaining the centroids of these posting lists in memory using a graph index. Upon a query, SPANN traverses the in-memory graph to identify top-$m$ nearest posting lists and loads them to host main memory. Then, it finds the top-$k$ ($k < m$) nearest neighbors within these $m$ posting lists via distance calculations. 

Although SPANN achieves low latency comparable to in-memory ANNS approaches, we find that its throughput for concurrent queries is quite limited. 
As shown in Figure~\ref{throughput}, SPANN achieves the peak QPS using only four CPU threads, and its throughput can not scale with more threads. We count the query latency in two stages: (\textit{i}) graph traversal in memory, and (\textit{ii}) processing posting lists from SSD. Figure~\ref{latency} shows that the query latency increases almost linearly with the number of threads. However, the latency of graph traversal almost remains stable, whereas the latency of processing posting lists increases significantly with the number of threads. The reason is that multiple queries concurrently read many and large-size posting lists from SSDs, resulting in severe I/O contention and high latency.



\subsection{Product Quantization}
To reduce the size of vector indices and computational costs for large-scale datasets, \textit{product quantization} (PQ)~\cite{IVF-PQ} has been explored recently for compressing high-dimensional vectors. Assume a dataset containing $N$ vectors is compressed with PQ, these vectors are first divided evenly into $M$ sub-spaces, and each contains $N$ sub-vectors. Then, these sub-vectors are clustered to generate a codebook, which contains a set of centroids of all clusters. The codebook allows each sub-vector to be approximated by its nearest centroid. The number of clusters per sub-space is typically set to 256, allowing each cluster ID to be represented by one byte. Once all codebooks are generated, each vector can be compressed into an $M$-byte PQ code. Upon a query, a distance lookup table is first generated, including all distances between a sub-query-vector and centroids per sub-space. Then, the approximate distance between the query vector $q$ and a compressed vector $v$ can be formulated as:
\begin{equation}
\widehat{dist}(q,v) = {\textstyle \sum_{i=1}^{M}} dist(q_{i} ,c_{i}(v_{i} ) ) 
\end{equation}
where $M$ denotes the total number of sub-spaces, \(q_{i}\) denotes the $i$-th sub-query-vector, and \(c_{i}(v_{i} )\) denotes the centroid of the $i$-th sub-space of the compressed vector. Thus, the distance between \(q_{i}\) and \(c_{i}(v_{i} )\)  can be easily retrieved by looking up the distance table using the PQ code as the address. Finally, the distance between $q$ and $v$ can be summed up with all $dist(q_{i} ,c_{i}(v_{i}))$. 

Essentially, PQ converts a distance calculation between vectors into multiple memory access operations, and thus poses a significant challenge for traditional CPU-based computing architectures due to the relatively high latency of DRAM accesses. Therefore, the PQ is usually accelerated by GPUs~\cite{JUNO,IVF-PQ} because it can fully utilize their high bandwidth memory to improve the query performance.

\subsection{Main Idea and Challenges}
\label{SPANN variant}
To circumvent the challenges of  substantial   computing and memory resource  requirements posed by billion-scale datasets,  our goal is to design a high-throughput, low-latency, cost-efficient, and high-accuracy ANNS system using SSDs and an entry-level GPU. However, a significant challenge for designing a GPU-accelerated ANNS system is that the limited capacity of GPU's HBM causes extensive data swapping between GPU and CPUs, significantly degrading the ANNS performance for large-scale datasets. 


\textbf{A Straightforward Solution using PQ and HI.} Fortunately, the PQ technique can significantly reduce the memory footprint of vectors, thereby alleviating the performance bottleneck associated with data transmission between CPUs and GPUs. As a result, PQ has the potential to fully harness the capabilities of GPUs to accelerate distance calculations involved in ANNS~\cite{Faiss_GPU}. Here, we first discuss a straightforward GPU-accelerated ANNS solution using PQ and \textit{hierarchical indexing} (HI) techniques. Except that all vectors are compressed using PQ, this straightforward solution uses the same hierarchical indices as SPANN. Upon a query, the ANNS engine first traverses the navigation graph to identify top-$m$ nearest posting lists, and  loads these compressed posting lists to the GPU's HBM for distance calculations. Then,  GPU finds the top-$n$ candidate vectors by calculating the distance between the query vector and each compressed vector in the top-$m$ nearest posting lists. Since PQ has a negative impact on the query accuracy,    these intermediate results obtained by the GPU should be re-ranked to improve the query accuracy.   During re-ranking, the raw data of the top-$n$ candidate vectors should be compared with the query vector to find the final top-$k$ nearest neighbours. 

\begin{figure*} [t] 
      \centering  
      \subfloat[Latency per query]{  
        \includegraphics[clip, width=0.3\linewidth]{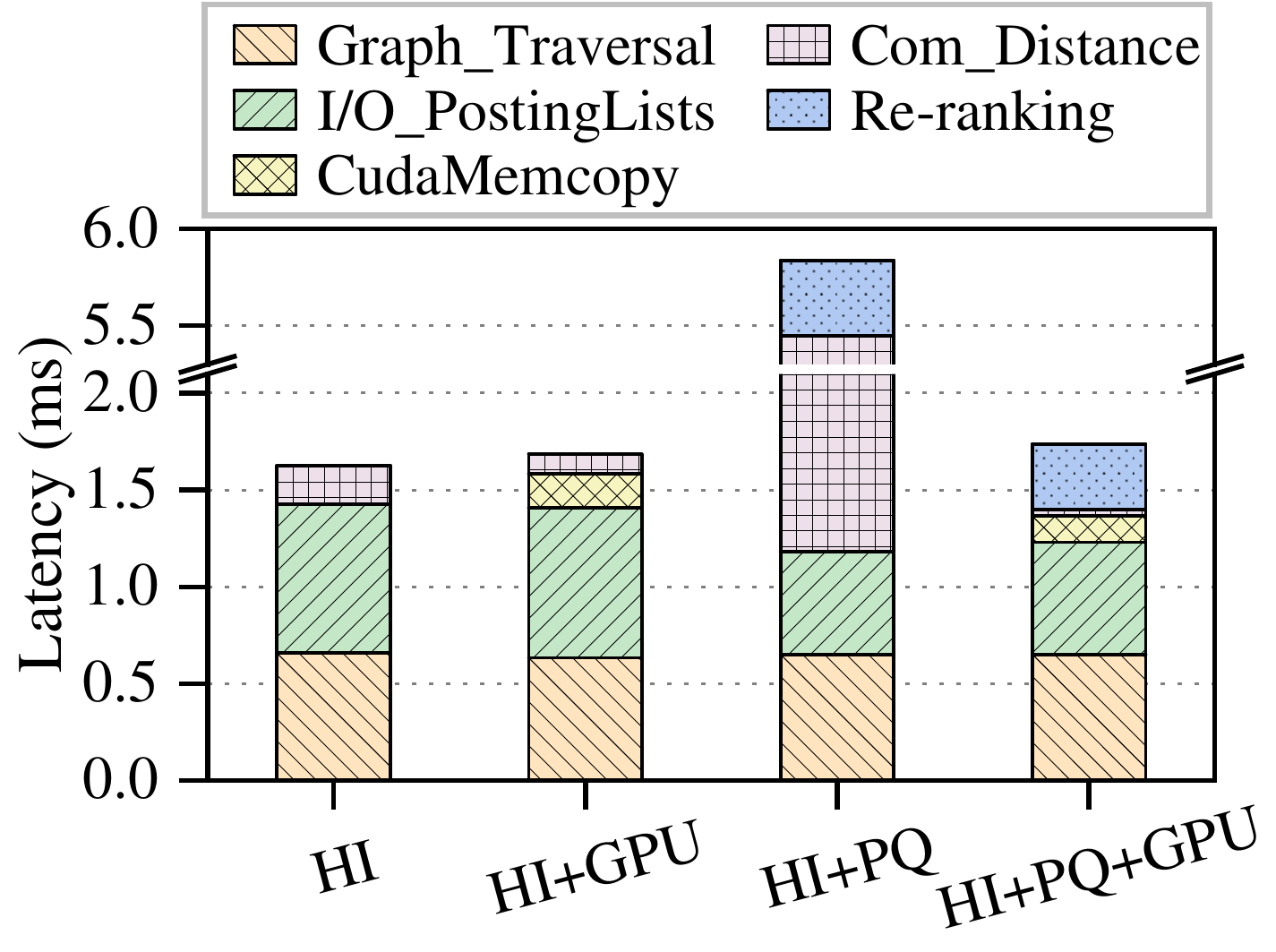}  
        \label{chutian1}  
      }  
      \subfloat[Throughput]{  
        \includegraphics[clip, width=0.225\linewidth]{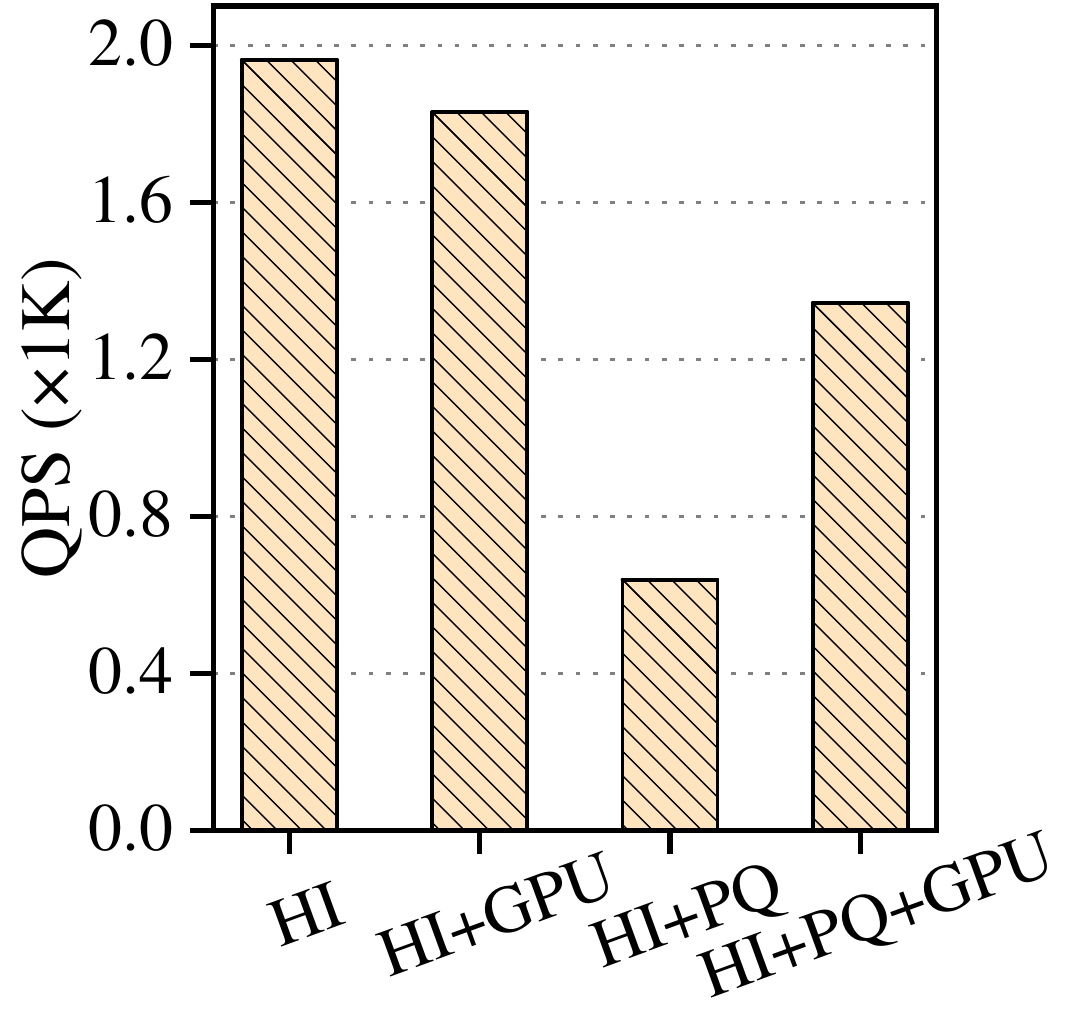}  
        \label{chutian2}  
      } 
      \subfloat[I/O numbers per query]{  
        \includegraphics[clip, width=0.225\linewidth]{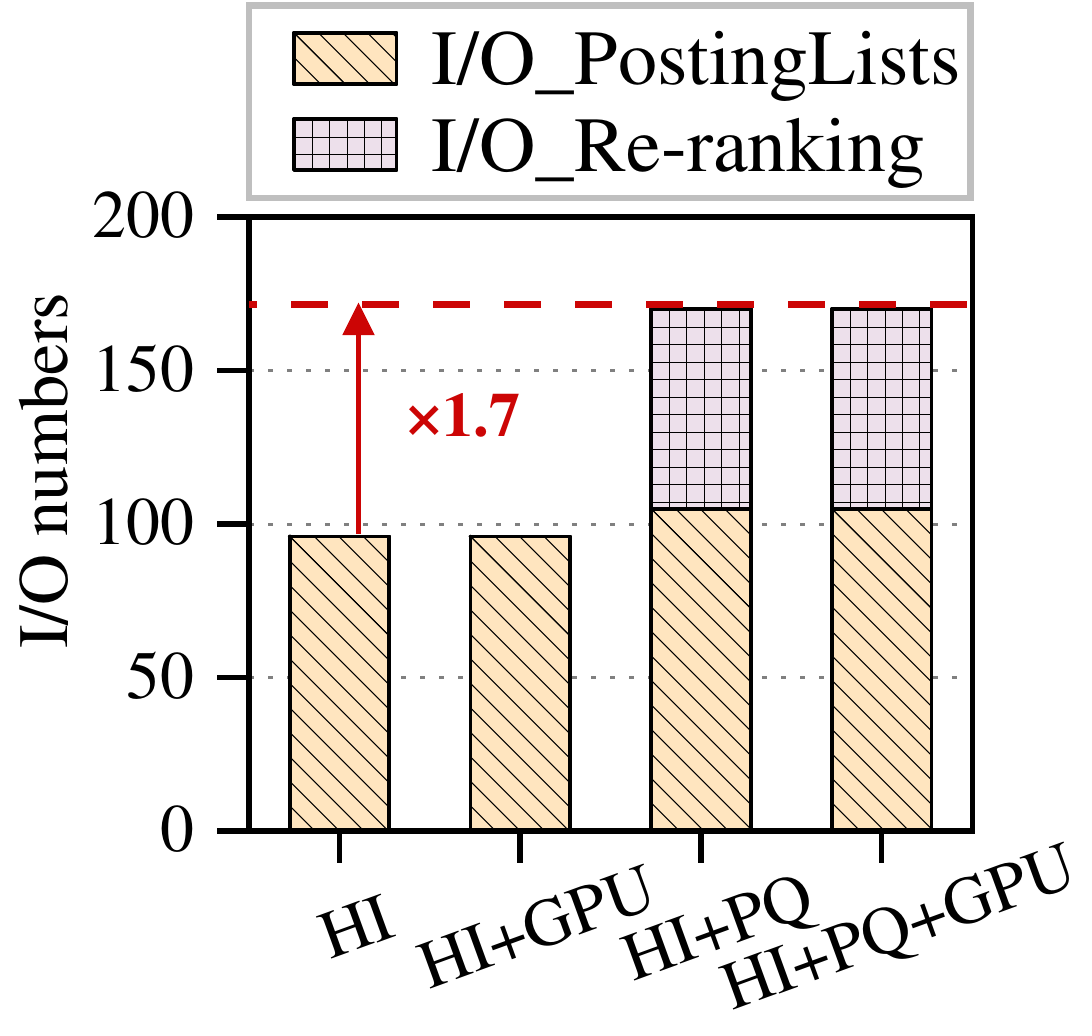}  
        \label{chutian3}  
      }  
      \subfloat[Data transmission per query]{  
        \includegraphics[clip, width=0.22\linewidth]{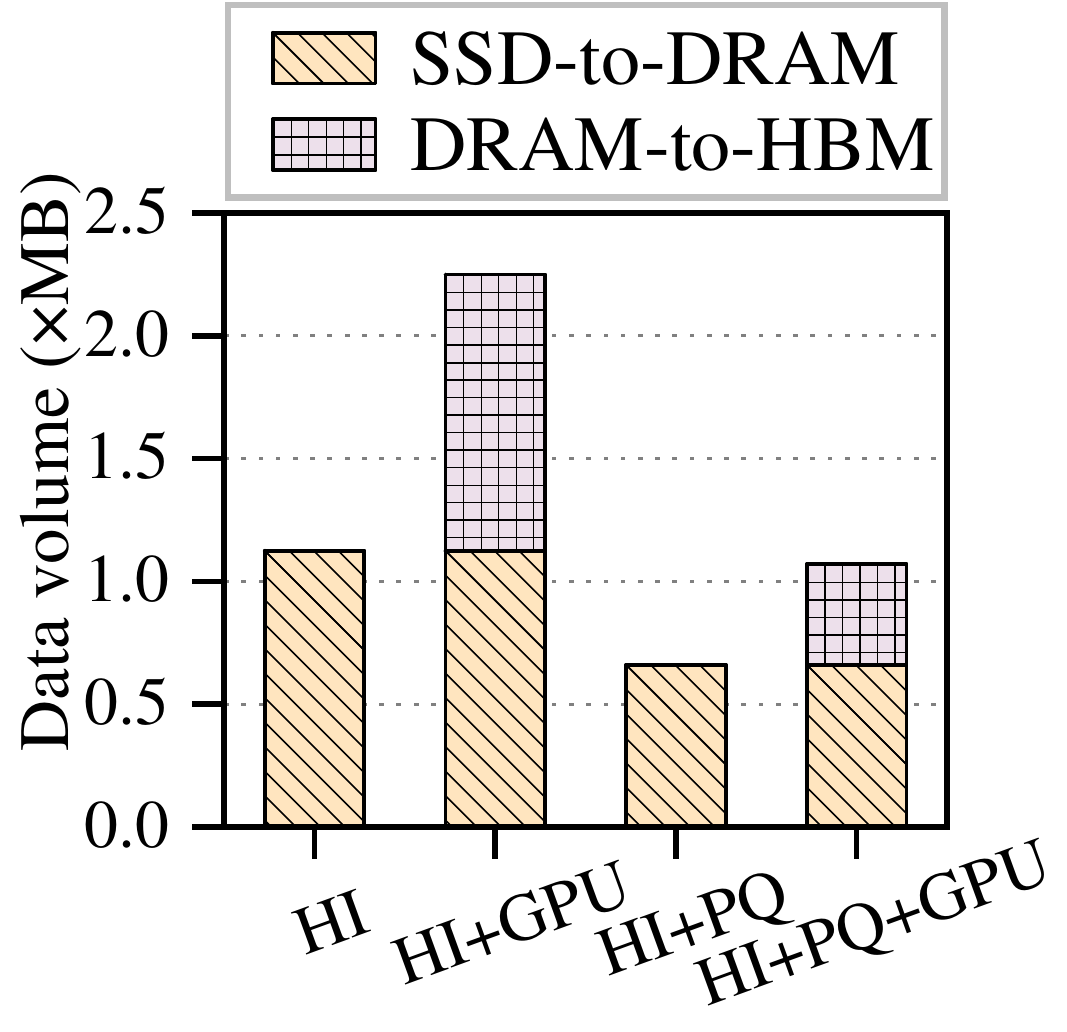}  
        \label{chutian4}  
      } 
      \vspace{-1ex}  
      \caption{Three combinations of \textit{hierarchical indexing} (HI), \textit{product quantization} (PQ), and GPU acceleration  
      } 
      \label{main idea}
      \vspace{-3ex} 
      \end{figure*}

\textbf{Observations}. Disappointingly, we find that the above solution does not achieve expected high performance. To better understand the root causes, we conduct four different experiments to evaluate three combinations of HI, PQ, and GPU acceleration techniques. In all experiments, different ANNS systems have to meet the same level of query accuracy. 
As shown in Figure~\ref{chutian1}, neither  the PQ nor the GPU acceleration can reduce the end-to-end query latency compared with the HI proposed by SPANN. Although ``HI+GPU'' can significantly reduce the latency of distance calculations, the overhead of transferring posting lists between CPUs and the GPU (i.e., CudaMemcpy) offsets the benefits of the GPU acceleration. For ``HI+PQ'', it still uses CPUs to process PQ-based posting lists. Since vectors are compressed using PQ, the I/O latency due to loading PQ-based posting lists from SSDs to main memory is reduced. However, the CPU  faces a new challenge in calculating distances between the query vector and compressed vectors due to intensive memory accesses, resulting in a significant increase of the end-to-end query latency. For ``HI+PQ+GPU'', the latency of distance calculations is reduced to an extremely low level. However, the CudaMemcpy and the additional re-ranking process incur substantial overheads, offsetting the benefits of using GPU. 
Moreover, none of these combinations achieve higher throughput than the original SPANN using HI solely, as shown in Figure~\ref{chutian2}. Particularly, a direct adoption of PQ to SPANN   even significantly reduces its QPS by 65\%. 


      \begin{figure} [t] 
      \centering  
      \subfloat[Accuracy vs. re-ranking numbers]{  
        \includegraphics[clip, width=0.51\linewidth]{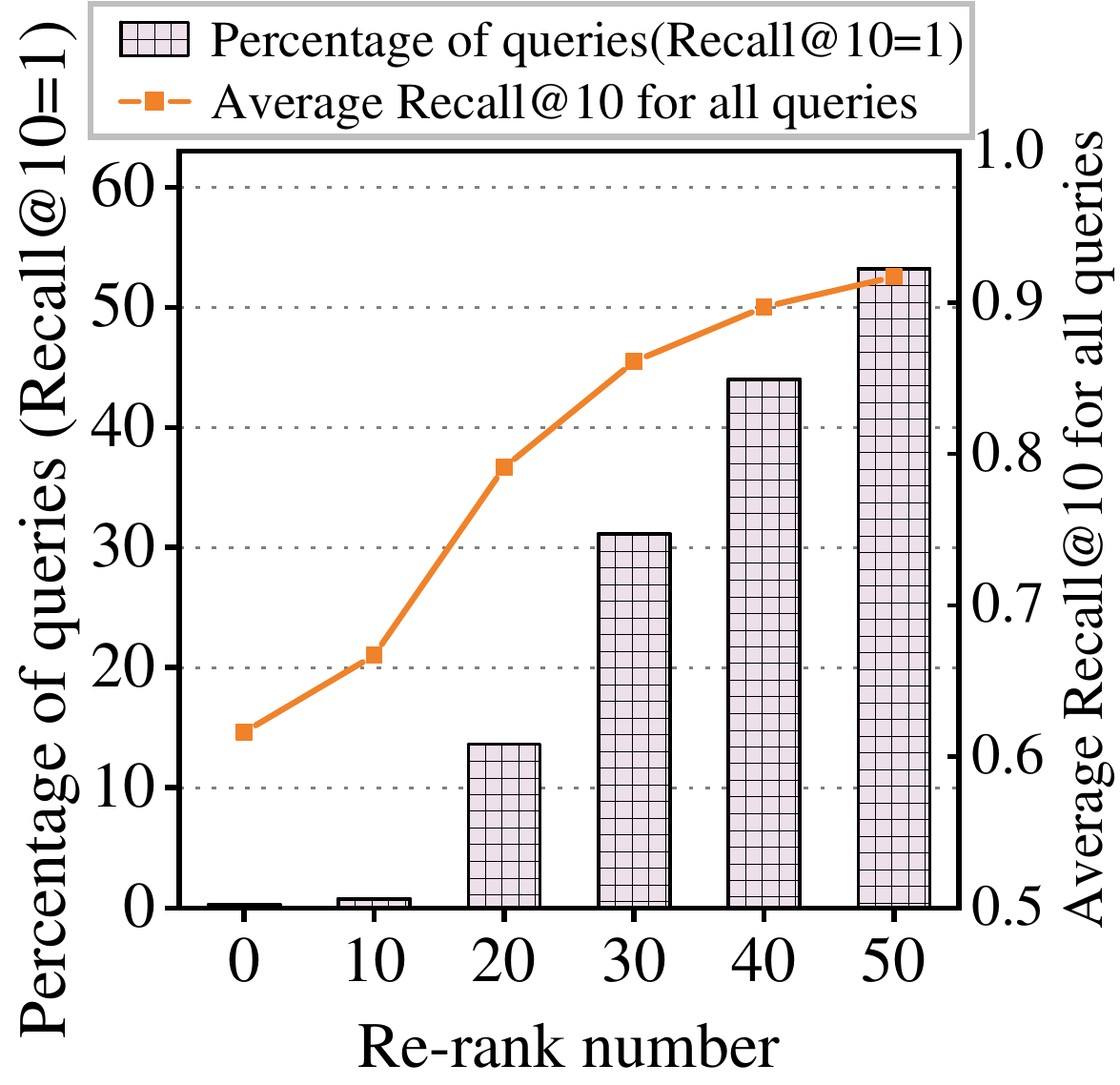}  
        \label{Querys_of_rerank_number}  
      }
      \subfloat[Re-ranking number per query]{  
        \includegraphics[clip, width=0.455\linewidth]{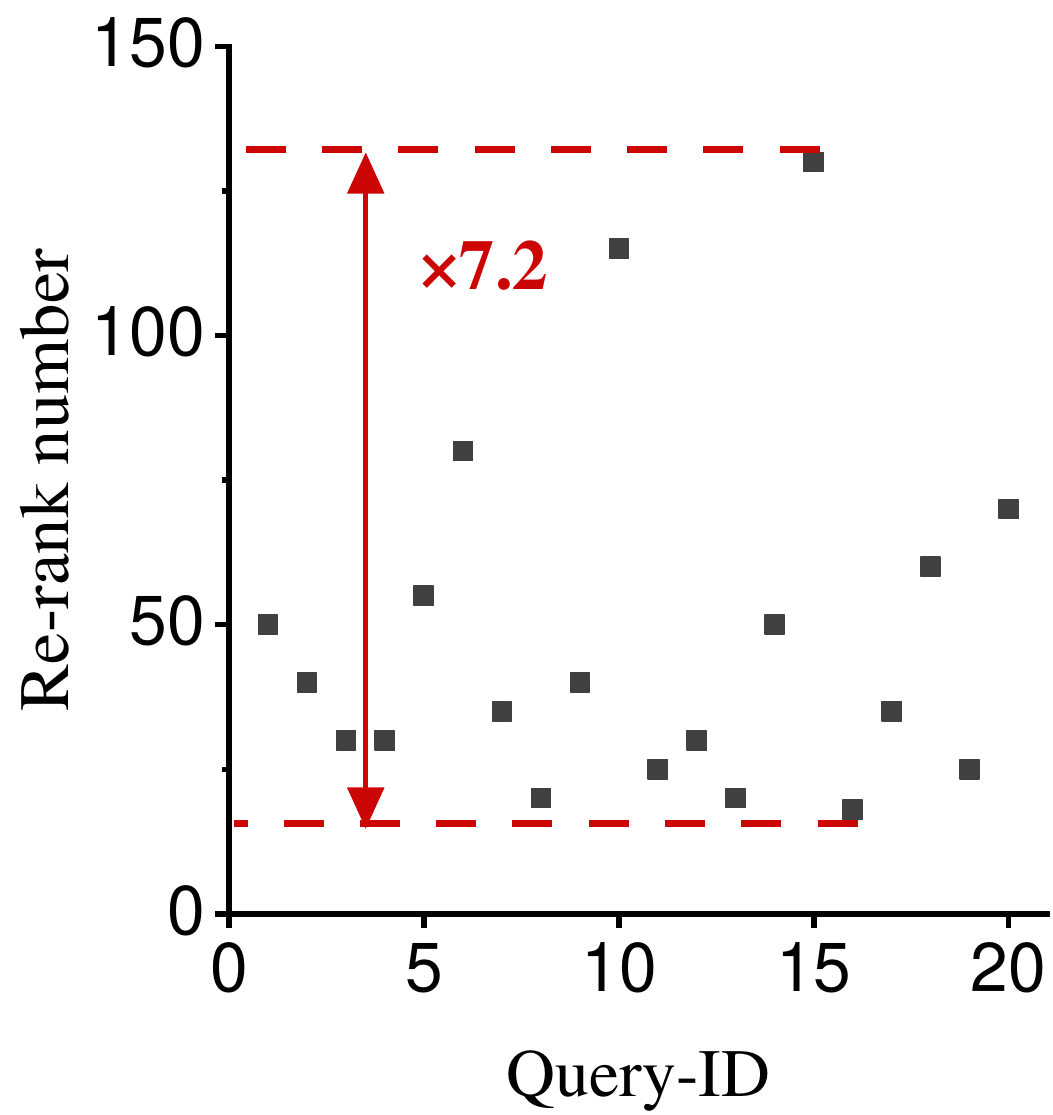}  
        \label{rerank_per_query}  
      }   
      \vspace{-1ex}  
      \caption{Differential characterization between queries
      } 
      \label{query}
      \vspace{-3ex} 
      \end{figure}

\textbf{Root Causes}. To reveal the root cause of such performance degradation, we measure the I/O numbers and the data volume transferred across SSDs, main memory and GPU's HBM required by each ANNS query on average. As shown in Figure~\ref{chutian3}, although the PQ technique significantly reduces the I/O size of posting lists from 12$\sim$48 KB to a page granularity (4 KB), it increases the number of I/Os by 70\% due to the re-ranking process. As a result, the I/O performance bottleneck shifts from the SSD's bandwidth to its \textit{input/output operations per second} (IOPS). Moreover, a large volume of posting lists are transferred  between CPUs and GPU, thereby offsetting the benefit of GPU acceleration, as shown in Figure~\ref{chutian4}.

\textbf{Challenge 1:} Overall, the combinations of HI, PQ, and GPU acceleration techniques cause even higher latency and lower throughput than SPANN that adopts HI solely. The root cause is that  the
GPU’s HBM still cannot accommodate all posting lists compressed by PQ, allowing extensive data transmission between GPU and
CPUs to become a new performance bottleneck. \textbf{Without a sophisticated design of the data layout across different devices and a careful collaboration among these three techniques, it is impossible to fully realize the GPU's potential for ANNS acceleration}.


\textbf{Challenge 2:} To achieve the same level of query accuracy, a re-ranking process is usually required to refine intermediate results generated by the GPU. The number of the top-$n$ vectors that should be re-ranked (i.e., the re-ranking number) is usually several times larger than the final top-$k$ nearest neighbors. To evaluate the impact of the re-ranking number on the query accuracy, we execute 10,000 queries by linearly increasing the re-ranking number.  As shown in Figure~\ref{Querys_of_rerank_number}, when  the re-ranking number is set to 40, about 42\% of queries have found the accurate top-10 nearest neighbors under $Recall@10=1.0$, while all queries achieve an accuracy level of $Recall@10=0.9$ on average. In this case,  it is only beneficial to increase the re-ranking number for straggler queries. 
Moreover, we find that  the minimum re-ranking numbers are very distinct for different ANNS queries, as shown in Figure~\ref{rerank_per_query}. This significant variance usually causes  unnecessary I/O operations and distance calculations if the number of re-ranked vectors is fixed for all queries.
However, \textbf{it is challenging to determine the minimum re-ranking number for each query under a given accuracy constrain}.

\textbf{Challenge 3:} The re-ranking process  introduces a number of I/O requests to  raw vectors on SSDs.  The size of a raw vector generally ranges from 128 bytes to 384 bytes, while the smallest operating unit 
of modern NVMe SSDs is typically a page (4 KB). \textbf{This mismatch in granularity often causes significant read amplification, resulting in extremely low I/O efficiency during re-ranking.} Fortunately, we find these vectors requiring re-ranking usually are highly similar to each other. This similarity offers an opportunity to mitigate the read amplification by carefully organizing the data layout on SSDs.

\begin{figure*}[t]
  \centering
  \includegraphics[clip,width=\linewidth]{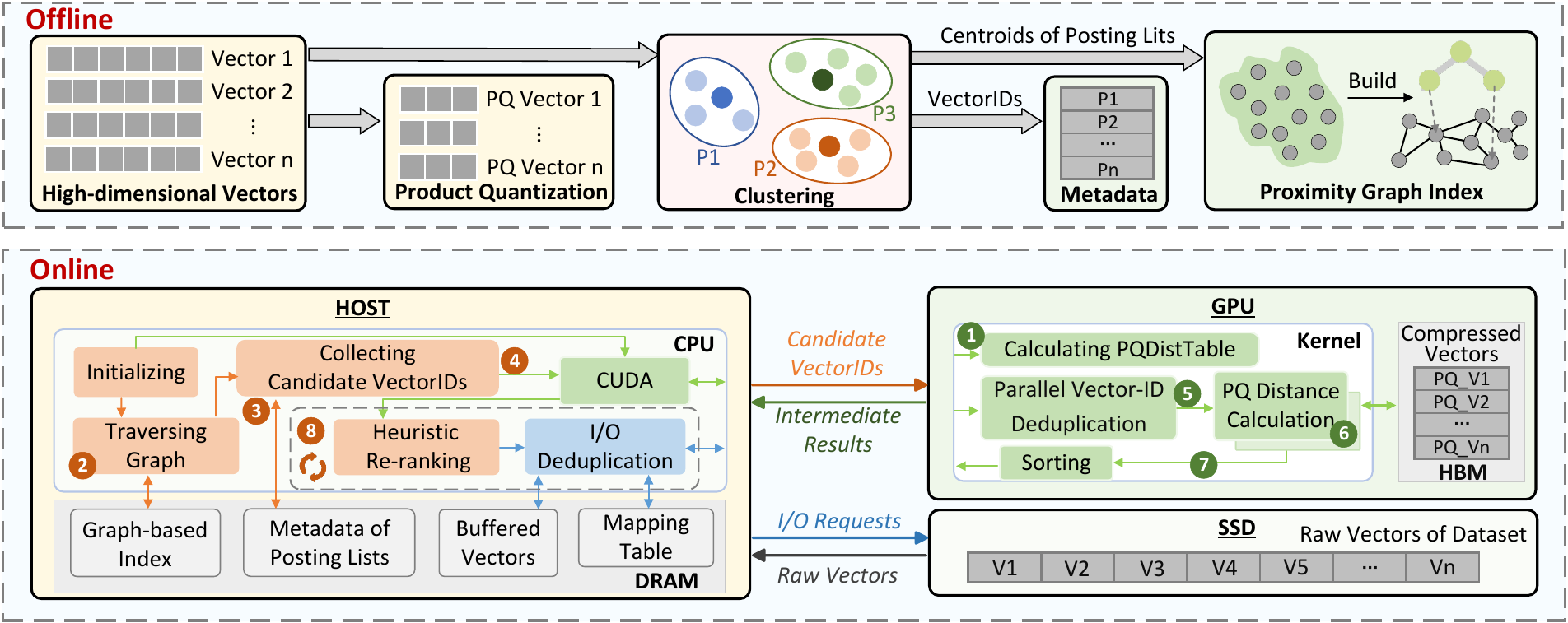}
  \caption{FusionANNS architecture}
  \label{overview}
\end{figure*}

\section{FusionANNS Overview}
We propose a multi-tiered index structure to enable CPU/GPU collaborative vector filtering and re-ranking. Figure~\ref{overview} shows an overview of the FusionANNS architecture.

\textbf{Offline Processing:}
Like most ANNS systems, FusionANNS also constructs multi-tiered indices in an offline manner. At first, FusionANNS  
exploits a clustering algorithm to partition the dataset into several posting lists. Then, a graph index is constructed using the centroids of these posting lists. After that, FusionANNS extracts the vector-IDs of each posting list as metadata, which is maintained in main memory along with the navigation graph. Finally, the intermediate posting lists are discarded, and the raw vectors and the PQ-based compressed vectors are stored on SSDs and GPU HBM, respectively. 

\textbf{Online Processing:}
Upon a vector query, FusionANNS first utilizes the GPU to generate the query vector's distance table for subsequent PQ distance calculations (\textbf{\circled{1}}). Meanwhile, the CPU traverses the in-memory graph to identify the top-$m$ nearest posting lists (\textbf{\circled{2}}). Then, the CPU consults the metadata to collect vector-IDs within these candidate posting lists (\textbf{\circled{3}}). After that, the CPU transfers these vector-IDs to the GPU and invokes GPU kernels (\textbf{\circled{4}}) for further processing.

When the GPU receives vector-IDs, it first deduplicates them using a parallel hash module (\textbf{\circled{5}}). For each vector-ID, the GPU reads the corresponding compressed vector from HBM and computes the PQ distance between it and the query vector. In this step, the GPU allocates a thread for each dimension to access the corresponding value in the distance table. Then, a coordinator thread accumulates these values to count the PQ distance (\textbf{\circled{6}}) for each candidate vector. Subsequently, the GPU sorts all distances and returns the top-$n$ vectors' ID to the CPU (\textbf{\circled{7}}). Finally, the CPU re-ranks these vectors using a heuristic re-ranking mechanism (\textbf{\circled{8}}) and returns the final top-$k$ nearest neighbors. During this stage, a redundancy-aware I/O deduplication mechanism is used to identify duplicate I/Os.

\section{FusionANNS Design}
In this section, we present the design of FusionANNS. We elaborate the detailed construction of multi-tiered index, the heuristic re-rank mechanism, and the redundant-aware I/O deduplication. 

\subsection{Multi-tiered Indexing }

\begin{figure}[t]
	\centering
	\begin{minipage}{\linewidth}
		\centering
		\includegraphics[width=1\linewidth]{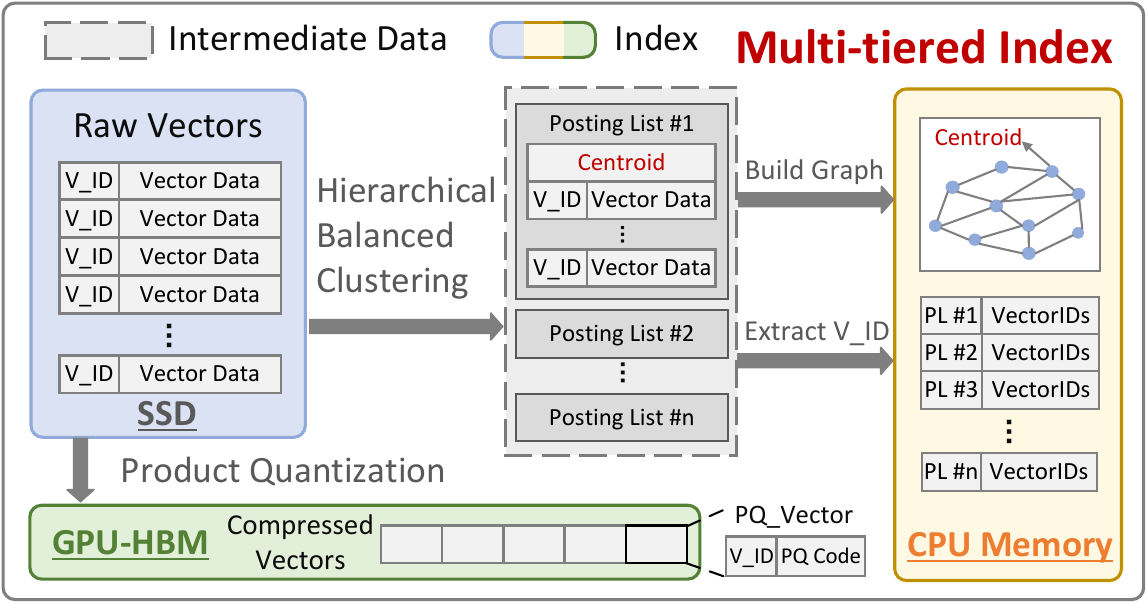}
		\caption{Mutil-tiered indices in FusionANNS}
		\label{index}
	\end{minipage}
 \vspace{-3ex}
\end{figure}

Figure~\ref{index} illustrates the structure of multi-tiered indices  resided in host main memory, GPU's HBM, and SSDs. We first use the hierarchical balanced clustering algorithm~\cite{cluster} to iteratively partition the dataset into several posting lists. Each posting list contains multiple vector-IDs and the corresponding vector content. The number of posting lists is only 10\% of the total number of vectors in a dataset. To improve the quality of each cluster, we use a replication mechanism to address the boundary concern~\cite{SPANN}. Specifically, when a vector lies on the boundary of multiple clusters, we assign this boundary vector to a cluster according to Equation~\ref{eq:rule1}.
\begin{equation}
v \in C_{i} \Leftrightarrow \text{Dist}(v, C_{i}) \le (1 + \epsilon) \times \text{Dist}(v,  C_1)
\label{eq:rule1}
\end{equation}
where \(v\)  represents the vector to be assigned, \(C_{i}\) denotes the $i$-th clusters.  Particularly, \(C_{1}\) represents the cluster that is closest to the vector \(v\). The parameter \(\epsilon\) determines the maximum distance in which a vector should be assigned simultaneously to multiple clusters. To balance the query accuracy and efficiency, each vector is assigned to eight clusters at most~\cite{SPANN}.

\textbf{In-memory Indices.} After the dataset is clustered, we build a graph index based on SPTAG~\cite{SPTAG}  using the centroids of all posting lists and store it in main memory. With this navigation graph, FusionANNS can efficiently identify the top-$m$ nearest posting lists for a query vector. This graph is constructed by continuously adding new vectors to an empty graph. When a vector is added as a new vertex, new edges are created to connects this newly-added vertex with its top-$k$ (typically
64) nearest neighbors. Then, its neighboring vertices should update their nearest neighbors to limit the maximum
number of edges. 
Unlike SPANN that stores all posting lists on SSDs, we extract only vector-IDs 
of each posting list as metadata (excluding vector content) and store it in memory, as shown in Figure~\ref{index}. When the graph index and metadata are generated, the intermediate posting lists can be discarded. 
Since the memory footprint of the graph and metadata is relatively small, FusionANNS can support billion-scale ANNS in a memory cost-efficient way using general-purpose servers.


\textbf{PQ-based Vectors in GPU's HBM}.  Since PQ can significantly reduce the memory footprint of high-dimensional vectors via lossy-compression, even an entry-level GPU such as NVIDIA V100 with 32 GB HBM   can accommodate all compressed vectors in its HBM for billion-scale datasets. In FusionANNS, we pin all compressed vectors in the HBM, avoiding extensive data swapping between GPUs and CPUs that is commonly experienced in previous GPU-accelerated ANNS systems~\cite{RUMMY,Faiss_GPU,GPU_Graph1}. Since in-memory indices still remain the benefit of the  replication mechanism for boundary vectors, FusionANNS can efficiently obtain all IDs of candidate vectors, and then sends these vector-IDs (excluding vectors' content) to the GPU for distance calculations. In this way, FusionANNS also eliminates the performance bottleneck caused by the limited PCIe bandwidth between CPUs and GPUs.  

\textbf{Raw Vectors on SSDs}. 
Unlike IVF-based SPANN that stores all posting lists on SSDs, FusionANNS only needs to store raw vectors on SSDs for re-ranking. Since the volume of raw vectors is almost 8 times smaller than that of posting lists, FusionANNS can significantly reduce the storage consumption. For each query, since only
the re-ranking process arises a few I/O
requests, FusionANNS can also  alleviate the I/O bottleneck of SSDs for concurrent queries. 

\textbf{Remarks.} Overall, our multi-tiered indexing approach is significantly different from previous hierarchical indexing techniques proposed by SSD-based or GPU-accelerated ANNS systems~\cite{SPANN,DiskANN,Faiss,RUMMY}. It can significantly reduce the storage footprints on HBM, main memory, and SSDs. It also significantly reduces the data volume transferred across SSDs, CPUs, and GPUs. This storage-saving and transmission-efficient approach effectively supports CPU/GPU collaborative filtering and re-ranking for fast and accurate ANNS.




\subsection{Heuristic Re-ranking}
\label{subsec:4.2}


Since PQ causes an accuracy loss during distance calculations, a re-ranking process is usually required to refine the intermediate results reported by the GPU. As mentioned in Section~\ref{SPANN variant},   to achieve the same level of query accuracy,  the minimum re-ranking numbers for different ANNS queries usually vary significantly.  Thus, a static configuration of the re-ranking number may cause unnecessary I/O operations and distance calculations, or result in an accuracy loss. 

To circumvent this problem, we propose a \textit{heuristic re-ranking} mechanism to minimize I/O operations and distance calculations.  The key idea is to set a relative large re-ranking number conservatively for high accuracy, and to terminate the re-ranking process immediately once the subsequent search is no longer beneficial for improving the query accuracy. To achieve this goal, we divide the re-ranking process into multiple mini-batches   and execute them sequentially. Each mini-batch contains the same number of candidate vectors.  Since all candidate vectors are sorted with their distances in ascending order, the mini-batch executed earlier usually has a higher possibility to identify more vectors that belong to the final top-$k$ nearest neighbours.  
Once a mini-batch is finished, we exploit a lightweight feedback control model to check whether subsequent mini-batches are beneficial for improving the query accuracy. 

To simplify the problem, we use a priority queue $Q$ (i.e., a max-heap) to maintain the current top-$k$ nearest neighbours. Initially, the max-heap is empty. For each mini-batch, we calculate the distances between the query vector and vectors within this mini-batch, and insert the vector whose distance is less than the current maximum distance in the max-heap. When a mini-batch is finished, we calculate the change rate of the max-heap according to Equation~\ref{eq:changerate}:
\begin{equation}
\Delta = \frac{|S_{n} - S_{n} \cap S_{n-1}|}{k}, 
\label{eq:changerate}
\end{equation}
\noindent where \( S_{n} \) and \( S_{n-1} \) represent the sets of vectors' IDs in the max-heap when the mini-batch $n$ and the mini-batch $n$-1 is just completed, respectively. $k$ represents the number of vectors maintained in the max-heap, i.e., the number of the final nearest neighbours.
We terminate the re-ranking process if the change rate of the max-heap for successive mini-batches is smaller than a given threshold $\epsilon$ continuously for total $\beta$ times.






\begin{algorithm}[t]
\renewcommand{\thealgocf}{1}
\caption{\label{alg:heuristic} Heuristic Re-ranking}
  \SetInd{0.1em}{1.5em}
    \KwIn{$Tasks,BatchSize,k,\epsilon, \beta$}
    \KwOut{$Q$}
    $Initialize\ Q \gets \textit{NULL}$
    
    $Initialize\ StabilityCounter \gets 0$
    
    \For{$i = 0; i \textless Tasks.size(); i += BatchSize$}
    {
    
        $S_{n-1} \gets Q.GetVectorIDs()$
        
        \For{$j = i; j \textless i + BatchSize; j ++$}
        {
        
            \tcc{Tasks[j] involves I/Os and computations}
            
            $Candidate\_Vector \gets GetDistance(Tasks[j])$
            
            $Q.insert(Candidate\_Vector)$
            
        }
        
        $S_{n} \gets Q.GetVectorIDs()$
        
        $\Delta \gets \frac{|S_{n} - S_{n} \cap S_{n-1}|}{k}$ \tcc{Calculating change rate $\Delta$}
        
        \eIf{$\Delta \leq \epsilon$}
        {
        
            $StabilityCounter \gets StabilityCounter + 1$

		\If{$StabilityCounter \geq \beta$}{

                \Return{Q} \tcc{Terminate re-ranking}
		}
        }
        {
            $StabilityCounter \gets 0$
        }
    }
    \Return{Q}
    \vspace{-1ex}
\end{algorithm}

        
        
    

Algorithm~\ref{alg:heuristic} presents the pseudo-code for the heuristic re-ranking. We first initialize the max-heap \textit{Q} as NULL, and use a \textit{StabilityCounter} to record the times that the change rate remains lower than $\epsilon$ continuously.  The size of \textit{Tasks} denotes the total number of vectors should be re-ranked in-batch. The parameter \textit{BatchSize} denotes the number of candidate vectors in a mini-batch. For each mini-batch, we retrieve the top-$k$ vectors' IDs from \textit{Q} before processing tasks in this mini-batch (line 4). Then, we sequentially perform each task including reading the raw vector from SSDs, calculating its distance to the query vector, and inserting this vector into \textit{Q} if its distance is less than the maximum value in the max-heap. When a mini-batch is finished, we collect the IDs of updated top-$k$ vectors from \textit{Q}, and calculate the change rate \( \Delta \) of \textit{Q} between these two successive mini-batches (line 9-10). If the change rate is lower than the given threshold \( \epsilon \), we increase the \textit{StabilityCounter} by one. Once the \textit{StabilityCounter} becomes larger than the given threshold \( \beta \), the re-ranking process is terminated. In contrast, if the change rate exceeds the threshold \( \epsilon \), we reset the \textit{StabilityCounter} and continue the following mini-batches. At last, we return the final top-$k$ vectors in \textit{Q}.  

\textbf{Remarks. } Our heuristic re-ranking algorithm can minimize I/O requests to SSDs and CPU resource consumption for distance calculation while guaranteeing high accuracy, and eventually reduces the latency of re-ranking.

\subsection{Redundant-aware I/O Deduplication}

FusionANNS uses raw vectors on SSDs to re-rank the intermediate results returned by the GPU. A straightforward approach is to store all raw vectors  sequentially on SSD pages. However, since a raw vector (128$\sim$384 bytes) is quite smaller than the page granularity (4KB), individual requests to these raw vectors often result in significant read amplification. Moreover, since the re-ranking process  introduces a lot of random and small I/O operations, the I/O latency has a crucial impact on the end-to-end query latency. Like most SSD-based ANNS systems~\cite{SPANN}, we adopt Direct I/O~\cite{Direct_1,Direct_2} to fully exploit the low-latency property of modern NVMe SSDs. 
To further improve I/O efficiency, we first optimize the data layout to improve the spatial locality on SSDs. Then, we exploit \textit{redundancy-aware I/O deduplication} to mitigate the effect of read amplification.

\begin{figure}[t]
	\centering
	\begin{minipage}{\linewidth}
		\centering
		\includegraphics[width=1\linewidth]{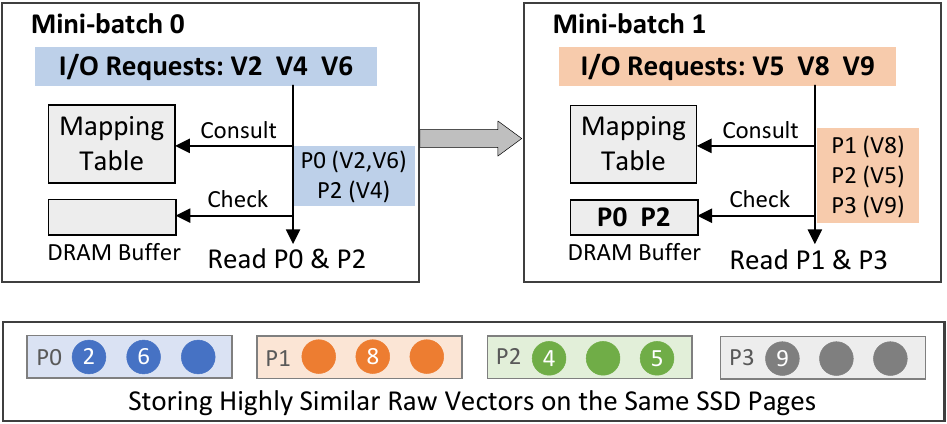}
		\caption{Optimized data layout and I/O deduplication}
		\label{Reordered Index}
	\end{minipage}
 \vspace{-3ex}
\end{figure}

\textbf{Optimized Storage Layout.} Although the vectors requiring re-ranking are obtained by PQ distances, they are all highly similar to the query vector, allowing them usually are spatially close to each other. This similarity offers an opportunity to mitigate the read amplification by carefully organizing the data layout on SSDs. Specifically, when the in-memory indexes are created offline, for each centroid in the navigation graph, we  use a bucket to store a number of raw vectors that are closest to the centroid. We note that there are not duplicate vectors among buckets. For each bucket, if it does not align with SSD pages, we combine buckets based on the size of unaligned portions using a max-min algorithm~\cite{max-min} to minimize the free space on a SSD page. Finally, We group all buckets as a single file and store it on SSDs, and use a table in memory to maintain the mappings between  vectors and SSD pages. 


\textbf{Intra- and Inter- Mini-batch I/O Deduplication.} Empowered by the optimized data layout, we design two I/O deduplication mechanisms, including merging I/Os mapped to the same SSD page within a mini-batch, and exploiting the DRAM buffer to eliminate redundant I/Os in subsequent mini-batches. Here, we use a simple example to describe
these mechanisms, as shown in Figure~\ref{Reordered Index}. Assume that there are two mini-batches in the re-ranking process, where the tasks of \textit{mini-batch 0} is to re-rank vectors: \textit{V2}, \textit{V4}, and \textit{V6}. The tasks of \textit{mini-batch 1} is to re-rank vectors:\textit{V5}, \textit{V8}, and \textit{V9}. When \textit{mini-batch 0} is executed, it first consults the mapping table to obtain the SSD page-IDs corresponding to the requested vectors. Since both \textit{V2} and \textit{V6} are stored in the same SSD page \textit{P0}, we can merge these two I/O requests and only read one SSD page to get \textit{V2} and \textit{V6}. Since \textit{P0} and \textit{P2} do not exist in the DRAM buffer, we directly read them to the DRAM buffer via two I/O requests. In \textit{mini-batch 1}, although \textit{V5}, \textit{V8}, and \textit{V9} are stored in different SSD pages, the DRAM buffer already contains \textit{P2} which includes \textit{V5}. Therefore, \textit{mini-batch 1} only needs to read \textit{P1} and \textit{P3} via two I/O requests.

\textbf{Remarks.} We optimize the data layout on SSDs to enable intra- and inter-Mini-batch  I/O deduplication mechanisms, which eventually mitigate the effect of read amplification and  improve I/O efficiency.



\section{Implementation}
We implement the system prototype of FusionANNS using 22K lines of codes in C++ and CUDA. FusionANNS can be widely deployed in general-purpose servers equipped with an entry-level GPU. 
Like most ANNS systems~\cite{SPFresh,SPANN,DiskANN,Starling }, we use each CPU thread to handle an individual query.


\textbf{Contention-free GPU Memory Management}. FusionANNS implements a GPU memory manager specifically for concurrent ANNS queries. During system initialization, we first load compressed vectors into GPU's HBM,  and use the remaining space as a memory pool. Then, we divide the memory pool into several independent blocks, each of which is assigned to a single query as working memory. Once a query is finished, its block can be assigned to other pending queries. This approach can avoid frequent memory allocations and lock contention between queries, improving the system performance.


\textbf{Efficient GPU Kernels}.
For each vector, we allocate multiple GPU threads according to their dimensions to calculate distances in parallel.
Moreover, we design a kernel to support parallel deduplication of vector-IDs using a hash algorithm. For a list of candidate Vector-IDs, we allocate a GPU thread for each vector-ID and use a spinlock to ensure that only one thread can access and update a hash table entry at a time. This approach can fully exploit  GPU's high parallelism to accelerate deduplication.

\section{Evaluation} \label{setup}



Our experiments are conducted on two servers, both using Ubuntu 22.04.4 LTS operating system. One is equipped with two Intel Xeon CPUs with 2.2 GHz 64 cores, 64 GB main memory, an entry-level NVIDIA V100 GPU with 32 GB HBM, and a Samsung 990Pro SSD with 2 TB storage capacity. 
This server is used to evaluate FusionANNS and other SSD-based ANNS solutions. Another server is equipped with the same CPUs and a GPU, but has 1 TB host memory. This large-memory server is used to evaluate a GPU-accelerated in-memory ANNS solution.

\textbf{Benchmarks.} In our experiments, we use three standard billion-scale datasets~\cite{BIGANN-Benchmarks} that are widely used by previous studies~\cite{DiskANN,SPANN,SPFresh}, as illustrated in Table~\ref{Datasets}.  Each benchmark simulates workloads using a set of query vectors. 

\textbf{Compared Solutions.} We compare FusionANNS with three representative ANNS solutions designed for billion-scale datasets, including two SSD-based solutions and a GPU-accelerated in-memory solution. We do not evaluate accelerator-based solutions using IVFPQ~\cite{IVF-PQ}  because its low accuracy can not meet the requirement of real-world applications, as reported in many previous studies~\cite{SPANN,DiskANN,HM-ANN,Starling}.
\begin{itemize}
    \item \textbf{SPANN}~\cite{SPANN} is the state-of-the-art \texttt{SSD-based ANNS  solution using the IVF index}. It is designed particularly for low latency. 
    \item \textbf{DiskANN}~\cite{DiskANN} is a \texttt{SSD-based ANNS  solution using the graph index}. It achieves high throughput, but suffers from extremely high latency.  
    \item \textbf{RUMMY}~\cite{RUMMY} is a state-of-the-art \texttt{GPU-accelerated in-memory ANNS solution using the IVF index}. It stores all vectors and their indices entirely in host memory. We extend RUMMY to support high-accuracy queries by adopting an advanced IVF index~\cite{SPANN}, without causing any performance degradation. 
\end{itemize}


      \begin{figure*} [t] 
      \centering  
      \subfloat[Throughput]{  
        \includegraphics[clip, width=0.42\linewidth]{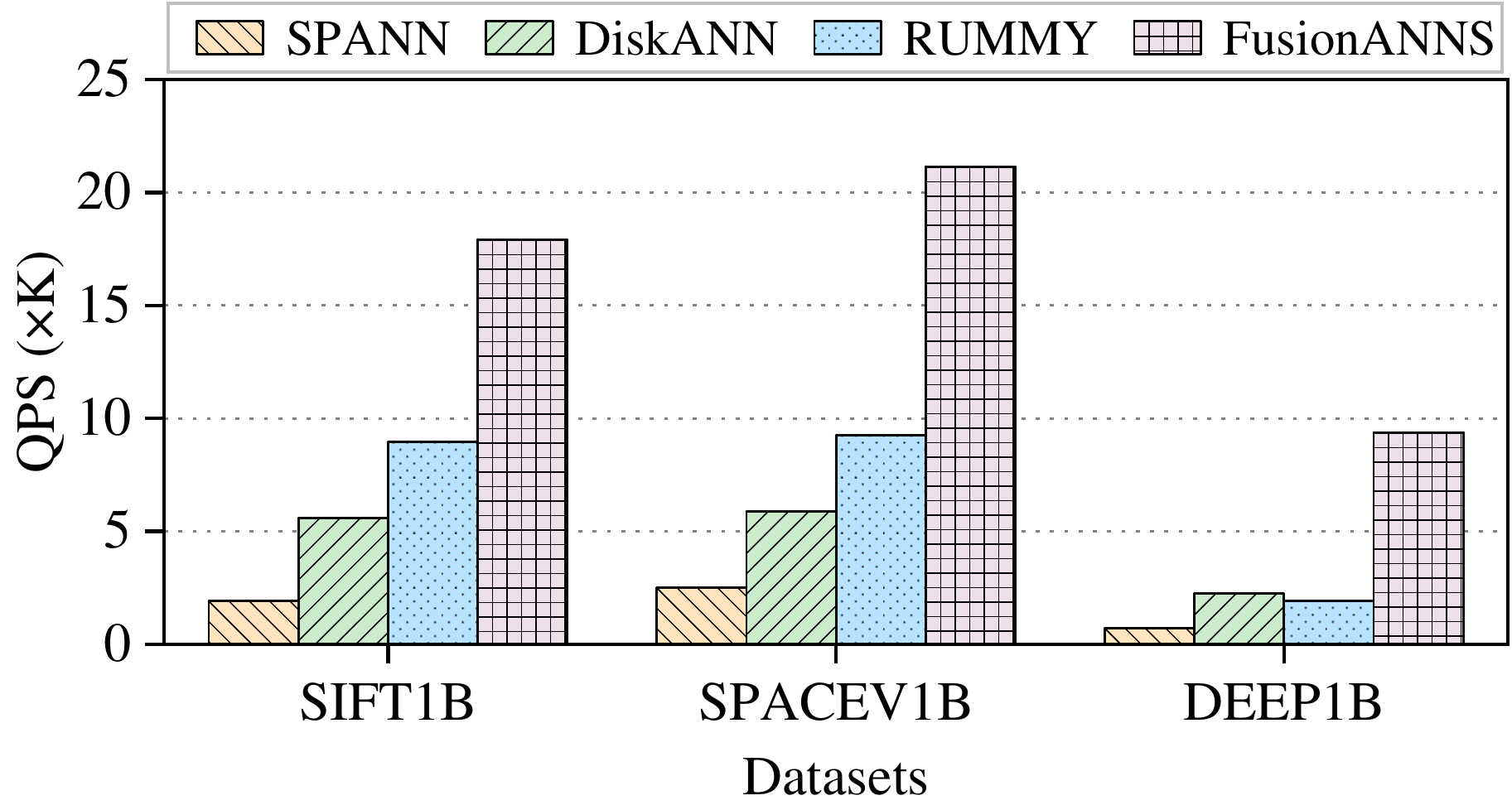}  
        \label{experment_QPS_datasets}  
      }\hspace{12mm}  
      \subfloat[Latency]{  
        \includegraphics[clip, width=0.42\linewidth]{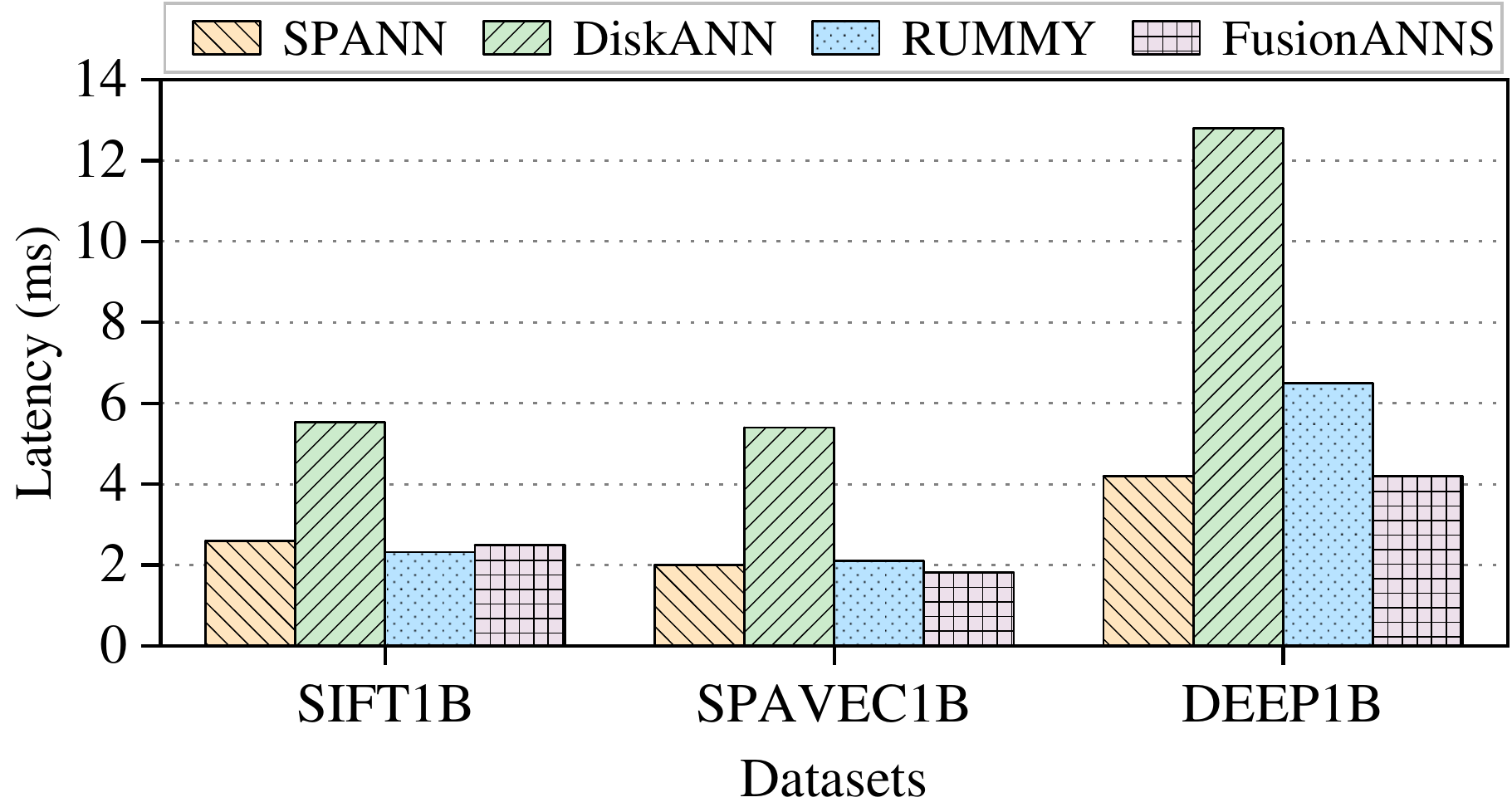}  
        \label{experment_latency_datasets}  
      } 
      \vspace{-1ex}  
      \caption{Throughput and latency of various ANNS systems using different datasets, under Recall@10=0.9
      } 
      \label{main idea}
      \vspace{-3ex} 
      \end{figure*} 

      \begin{figure*} [t] 
      \centering  
      \subfloat[Recall vs. QPS]{  
        \includegraphics[clip, width=0.42\linewidth]{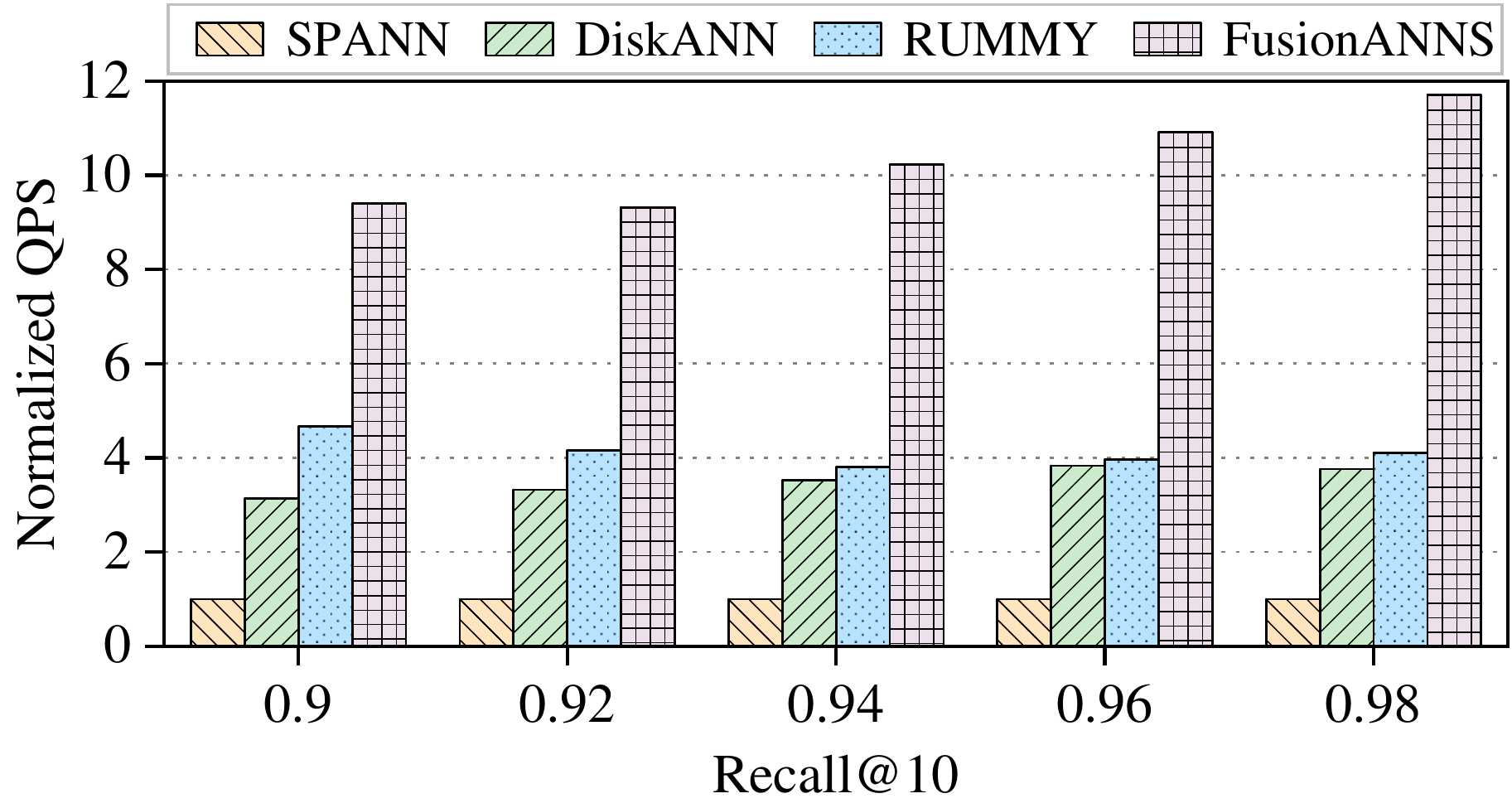}  
        \label{experment_QPS_recall}  
      }\hspace{12mm}  
      \subfloat[Recall vs. latency]{  
        \includegraphics[clip, width=0.42\linewidth]{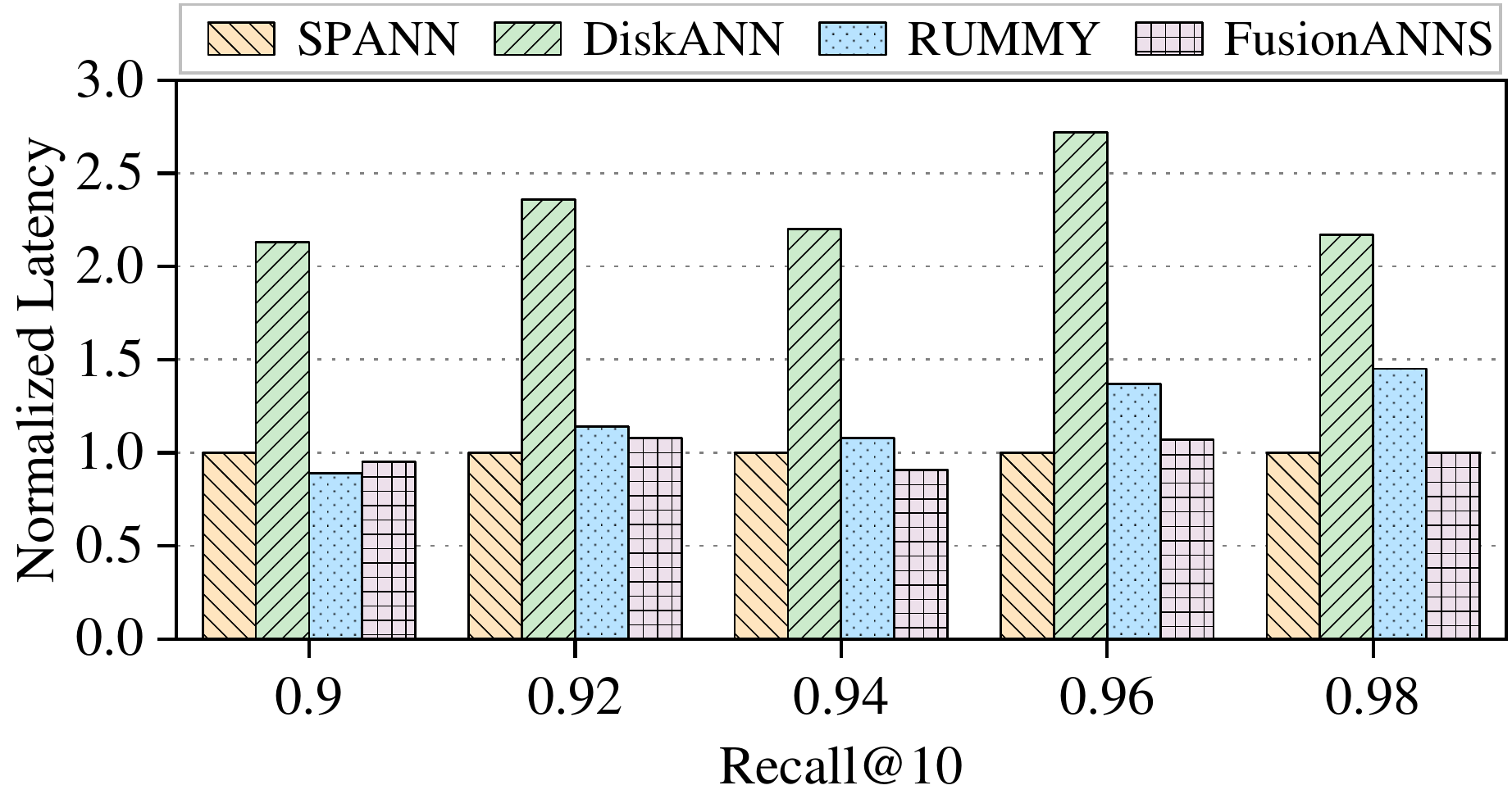}  
        \label{experment_latency_recall}  
      } 
      \vspace{-1ex}  
      \caption{Normalized throughput and latency of different ANNS systems using SIFT1B,  under different accuracy levels
      } 
      \label{main idea}
      \vspace{-3ex} 
      \end{figure*} 
\textbf{Performance Metrics.} We use \textit{ query per second} (QPS) and average latency to evaluate the performance of various ANNS systems. Like previous studies~\cite{SPFresh,CXL-ANNS,SmartANNS}, unless specified otherwise, the query accuracy is evaluated by Recall@10, which represents the proportion of the top-\( 10 \) results containing the ground-truth nearest neighbors for an ANNS query. To achieve a given accuracy level such as Recall@10=0.9, we can adjust two parameters in FusionANNS for different datasets, i.e., the number of top-$m$ nearest posting lists retrieved from the graph index, and the top-$n$ candidate vectors requiring a re-ranking process. 

\begin{table}[t]
\centering
\caption{Datasets (One Billion)}
\label{Datasets}
\footnotesize

\begin{tabular}{@{}ccccc@{}}
\toprule
\textbf{Dataset} & \textbf{Dimension} & \textbf{Raw Data Size} & \textbf{Data Type} & \textbf{Domain} \\ \midrule
SIFT1B           & 128                & 119 GB              & uint8           & Image              \\
SPACEV1B         & 100                & 93 GB               & int8           & Web Search         \\
DEEP1B           & 96                 & 358 GB              & float32          & Image              \\ \bottomrule
\end{tabular}
\end{table}
\subsection{Performance}
We measure the QPS and latency of different ANNS systems under the same constraint of query accuracy. For these experiments, we gradually increase the number of threads for concurrent queries till these systems achieve the peak QPS. 

\textbf{Performance under Different Datasets.} We compare FusionANNS with other ANNS systems using three datasets. Figure~\ref{experment_QPS_datasets} and Figure~\ref{experment_latency_datasets} show QPS and latency, respectively, under the accuracy level of  Recall@10=90\%. Compared with  SSD-based SPANN and DiskANN, FusionANNS can significantly improve  QPS by 9.4-13.1\(\times\) and 3.2-4.3\(\times\),  respectively. 
Although SPANN shows very low latency, its throughput is rather low compared with other ANNS systems. In contrast,
FusionANNS achieves low latency comparable to SPANN, but significantly improves the throughput.  
These results demonstrate that FusionANNS achieves both high throughput and low latency for these billion-scale datasets. Such improvement mainly stems from the multi-tiered index enabled CPU/GPU collaborative filtering and re-ranking  techniques. Because FusionANNS can  avoid extensive data swapping between CPUs and the GPU, and can also mitigate  unnecessary I/O operations on SSDs, it eliminates the I/O performance bottleneck due to limited PCIe bandwidth.



Compared with the GPU-accelerated in-memory solution-RUMMY, FusionANNS improves the QPS by 2-4.9\(\times\) while remaining low latency for different datasets. Notably, RUMMY exhibits much lower performance for the DEEP1B dataset compared with other datasets. The reason is that  the data transfer for a larger dataset from main memory to GPU's HBM consumes more PCIe bandwidth, making the bandwidth bottleneck between CPUs and GPU more pronounced. In contrast,  FusionANNS only needs to transfer lightweight vector-IDs rather than the vectors' content between CPUs and the GPU. Thus,  FusionANNS achieves significant performance improvement relative to RUMMY.

\textbf{Performance under Different Accuracy Levels.} To evaluate the performance of FusionANNS under different accuracy levels, we change the Recall@10 from 90\% to 98\%. Figure~\ref{experment_QPS_recall} and Figure~\ref{experment_latency_recall} shows  QPS and latency under different levels of accuracy using the SIFT1B dataset, respectively. All results are normalized to SPANN. FusionANNS achieves about 9.4-11.7~\(\times\) and 3.2~\(\times\) QPS improvement compared with SPANN and DiskANN, respectively. FusionANNS achieves more QPS improvement with the increase of query accuracy compared with SPANN, and  even offers much lower latency than the in-memory RUMMY under the constraint of higher accuracy. The root cause is that the CPU/GPU corroborative filtering mechanism  can effectively eliminate data swapping between main memory and GPU's HBM, an thus can extend the search space to meet a higher accuracy level whiling still remaining high performance. In contrast, other ANNS solutions cause more I/O operations and distance calculations when the search space becomes larger.




      \begin{figure*} [t] 
      \centering  
      \subfloat[SIFT1B]{  
        \includegraphics[clip, width=0.32\linewidth]{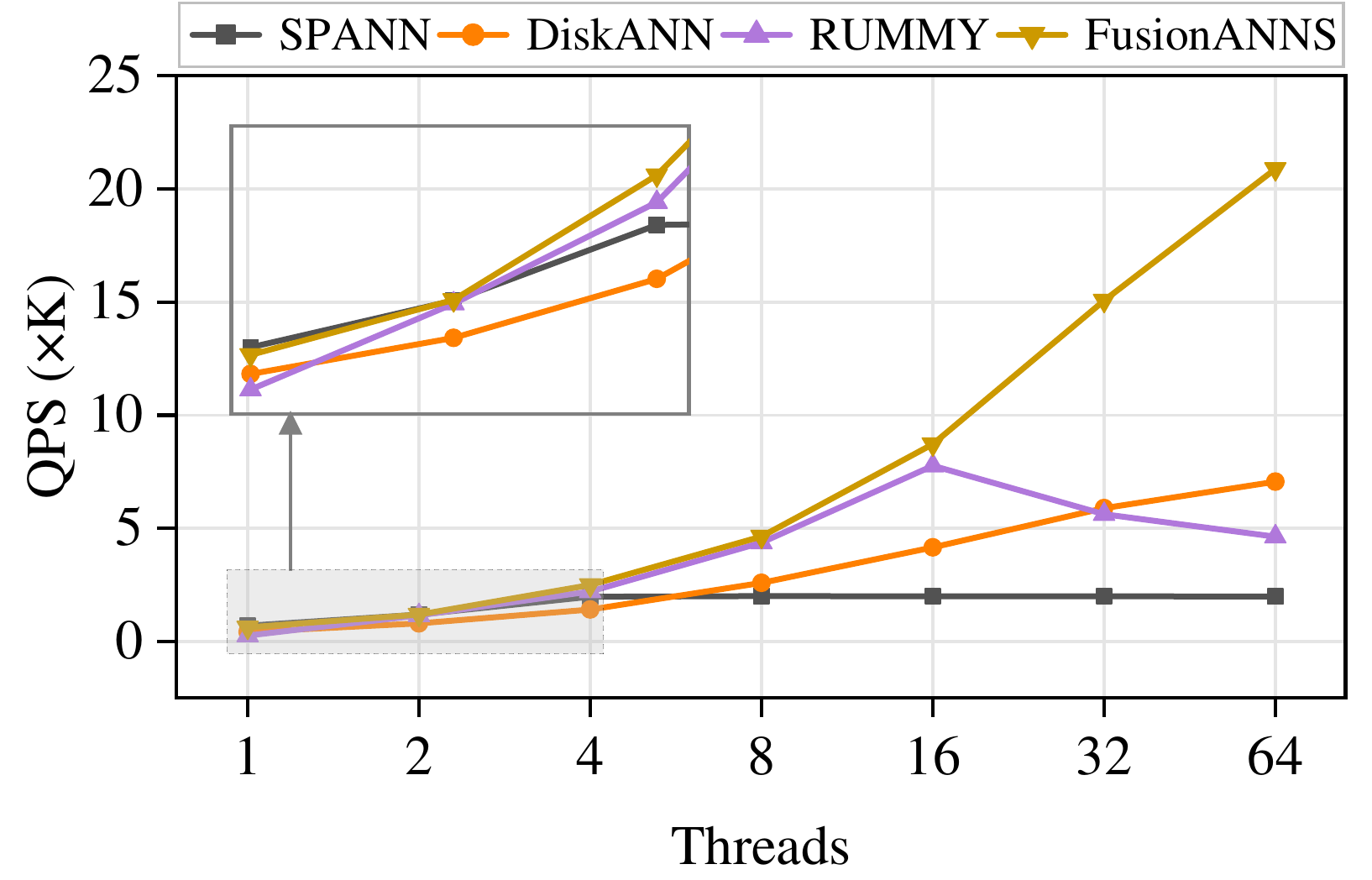}  
        \label{Scala:QPSSIFT}  
      }
      \hfill
      \subfloat[SPACEV1B]{  
        \includegraphics[clip, width=0.32\linewidth]{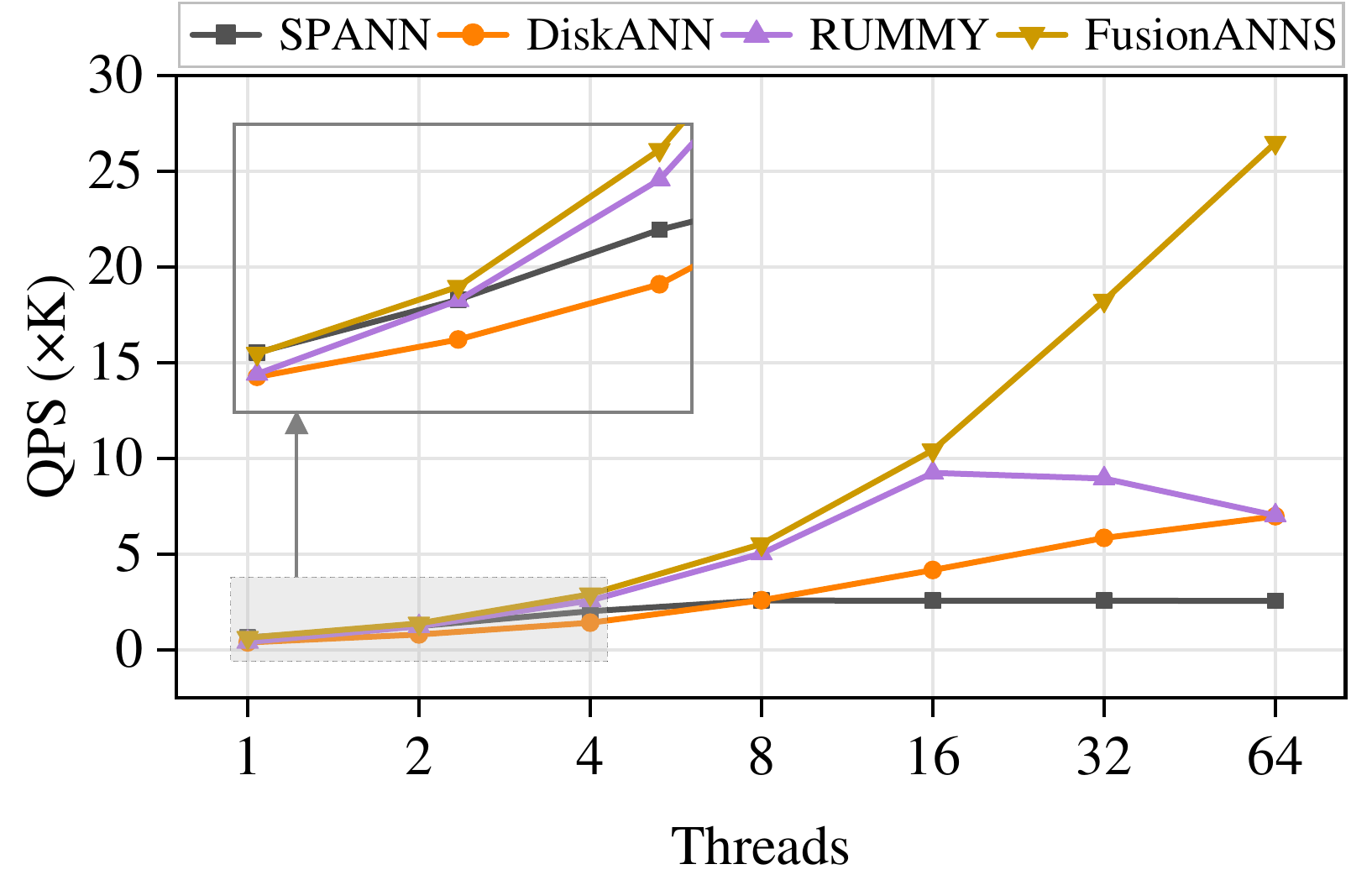}  
        \label{Scala:QPSSPACEV}  
      }
      \hfill
      \subfloat[DEEP1B]{  
        \includegraphics[clip, width=0.32\linewidth]{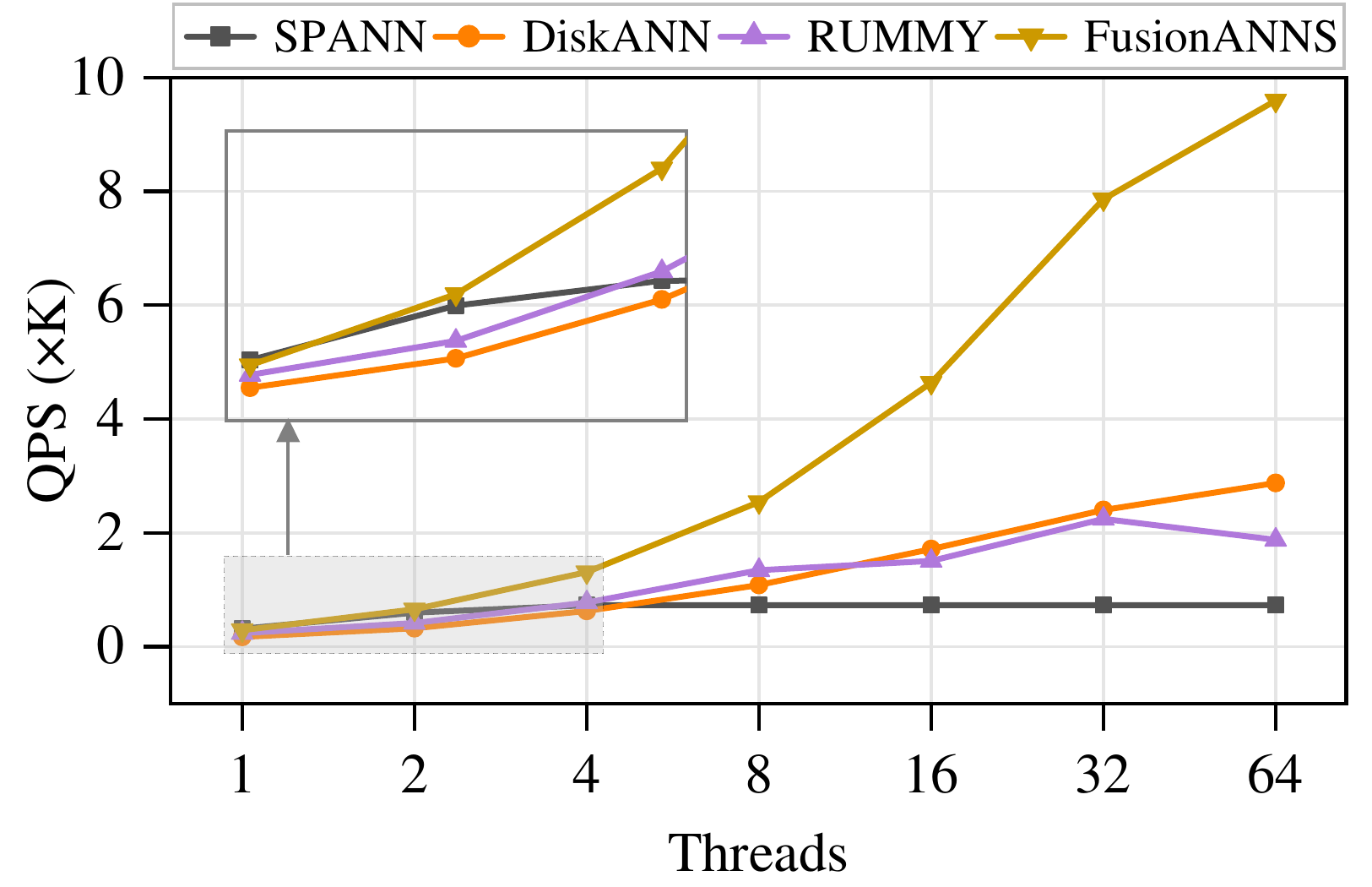}  
        \label{Scala:QPSDEEP}  
      }

      \subfloat[SIFT1B]{  
        \includegraphics[clip, width=0.32\linewidth]{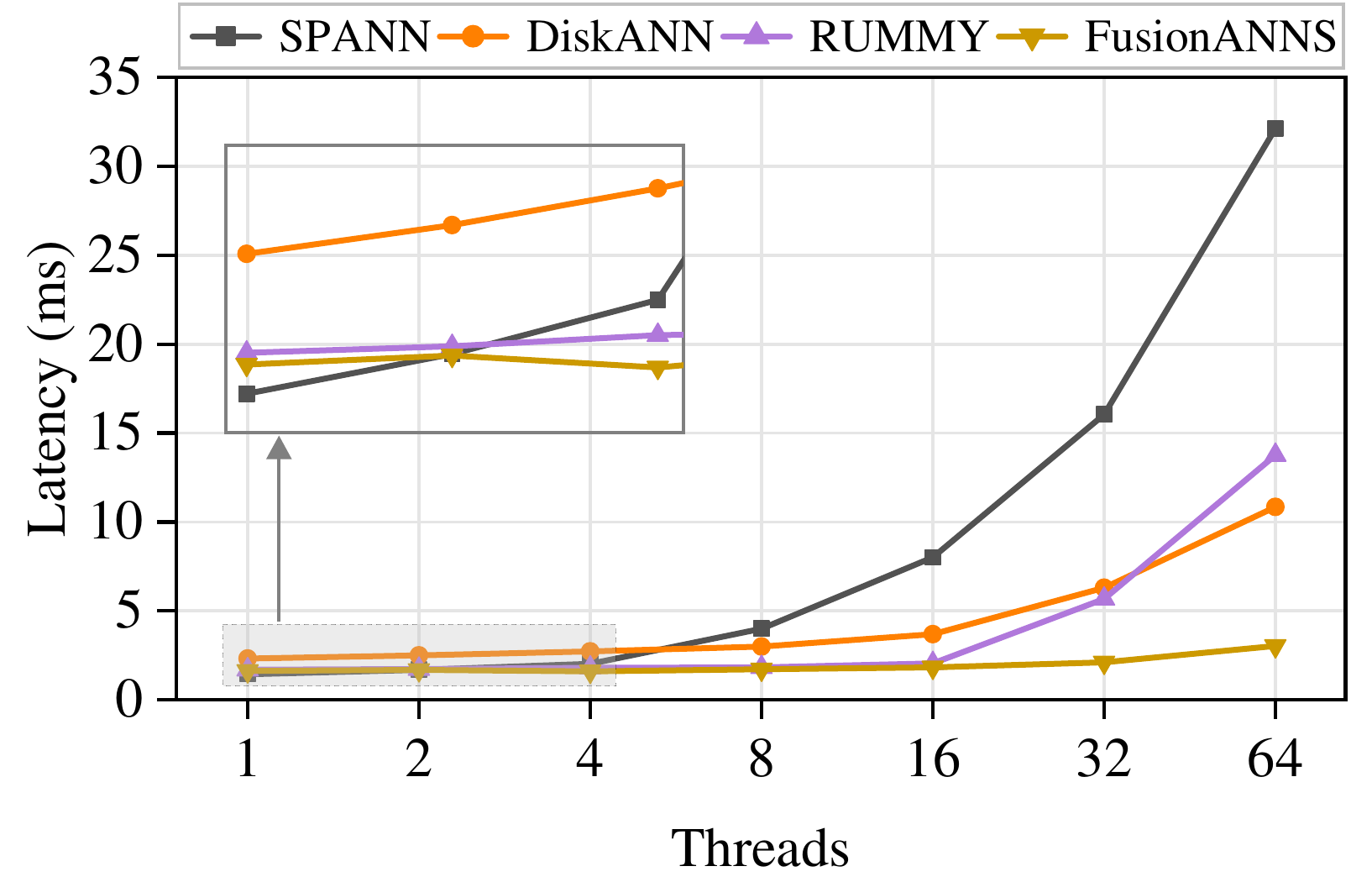}  
        \label{Scala:QPSSIFT}  
      }
      \hfill
      \subfloat[SPACEV1B]{  
        \includegraphics[clip, width=0.32\linewidth]{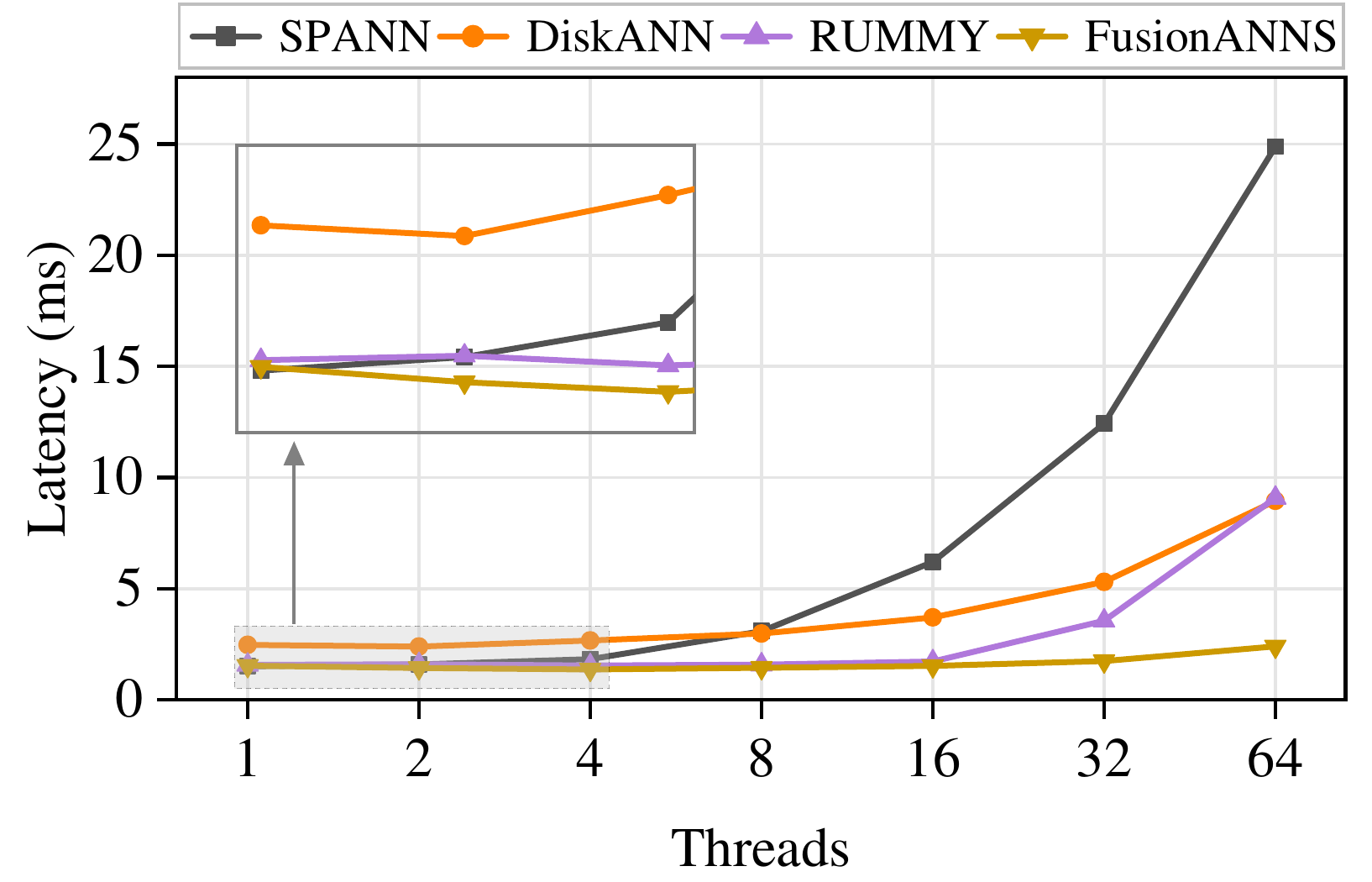}  
        \label{Scala:QPSSPACEV}  
      }
      \hfill
      \subfloat[DEEP1B]{  
        \includegraphics[clip, width=0.32\linewidth]{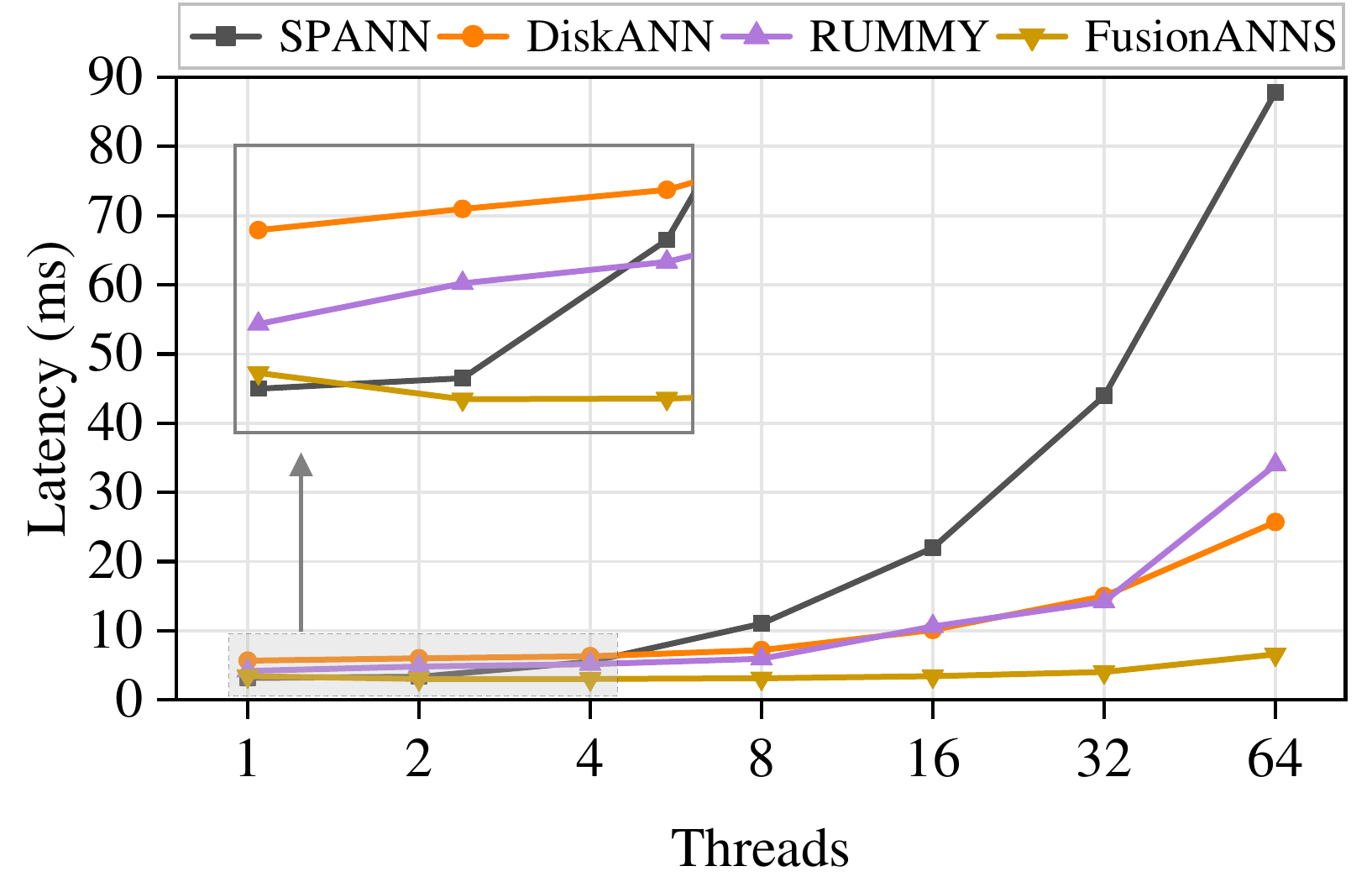}  
        \label{Scala:QPSDEEP}  
      } 
      \vspace{-1ex}  
      \caption{The throughput and latency of different ANNS systems vary with the number of threads.
      } 
      \label{Scalability}
      \vspace{-4ex} 
      \end{figure*} 

\subsection{Scalability}
The throughput of ANNS systems is highly correlated to the number of CPU threads. 
To evaluate the performance scalability of different ANNS systems, we increase the number of threads exponentially by a factor of 2.


Figure~\ref{Scalability} shows the QPS and latency of different ANNS systems using different numbers of threads. FusionANNS show a significant growth in QPS when the number of threads increases from 1 to 64. For all ANNS systems, the QPS is almost the same when only one thread is used. However, when the number of threads increases to 64, FusionANNS significantly improves QPS by up to 13.2~\(\times\), 3.8~\(\times\), and 5.1~\(\times\) while remaining low latency, compared with SPANN, DiskANN, and RUMMY, respectively. SPANN achieves its peak QPS using only 4 threads, and its latency increases significantly with more threads. 
Notably, for SIFT1B and SPACEV1B datasets, the QPS of RUMMY peaks at 16 threads, and then decreases distinctly. Meanwhile, the latency of RUMMY also significantly increases when the number of threads becomes larger than 16.   This reason of such limited scalability is that more concurrent queries arise a large amount of data transmission between CPUs and the GPU, which lead to significant PCIe bandwidth contention. Due to the bigger vector size of the DEEP1B dataset,
RUMMY suffers from more severe bandwidth contention, and thus its QPS is even lower than DiskANN, as shown in Figure~\ref{Scala:QPSDEEP}.
FusionANNS shows much better scalability than others even using limited memory resource because it eliminates the data swapping between CPUs and the GPU, and also significantly improves the I/O efficiency on SSDs.  



      \begin{figure*} [t] 
      \centering  
      \subfloat[Throughput]{  
        \includegraphics[clip, width=0.32\linewidth]{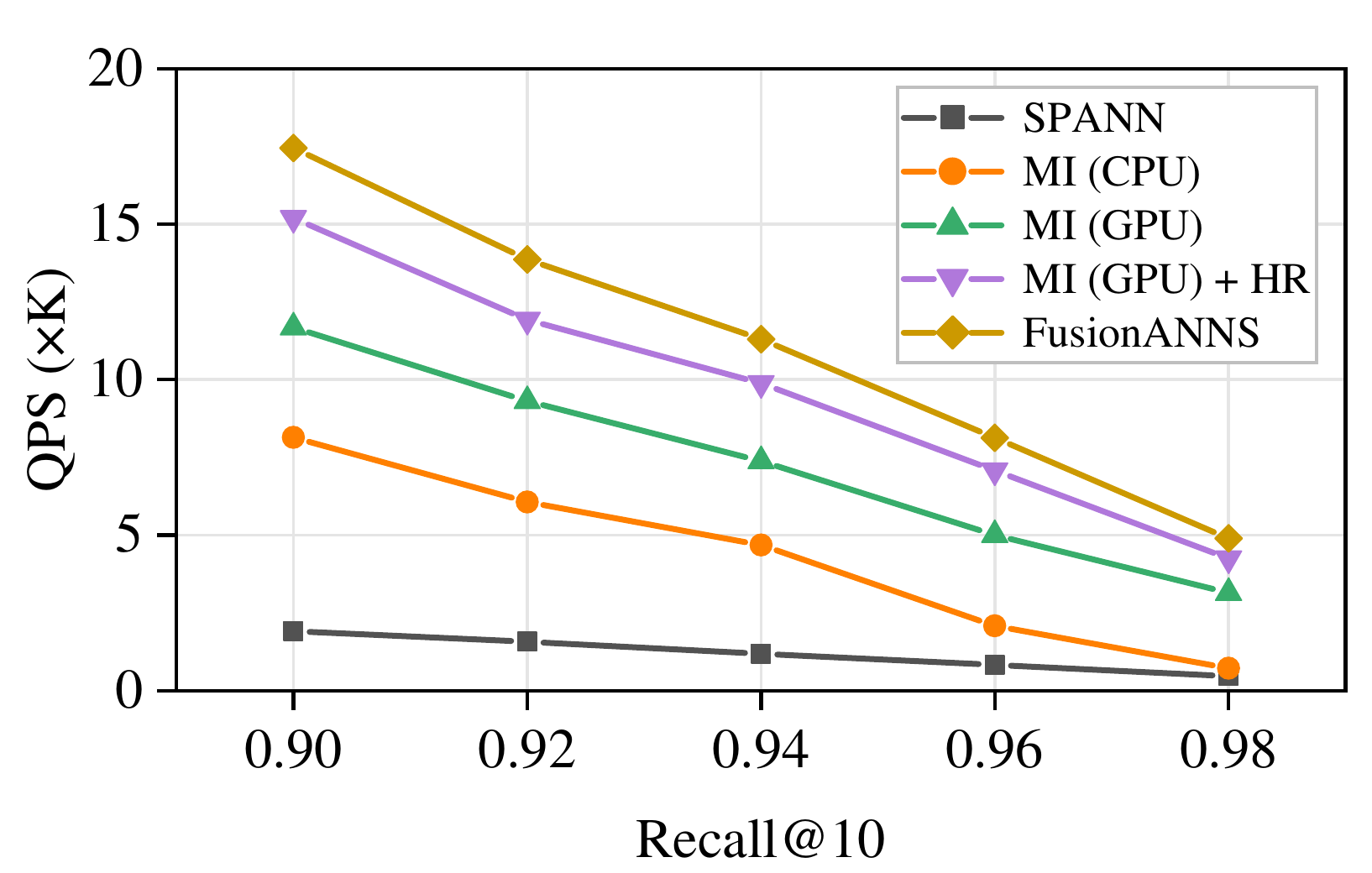}  
        \label{breakdownQPS}  
      }  
      \subfloat[Latency]{  
        \includegraphics[clip, width=0.32\linewidth]{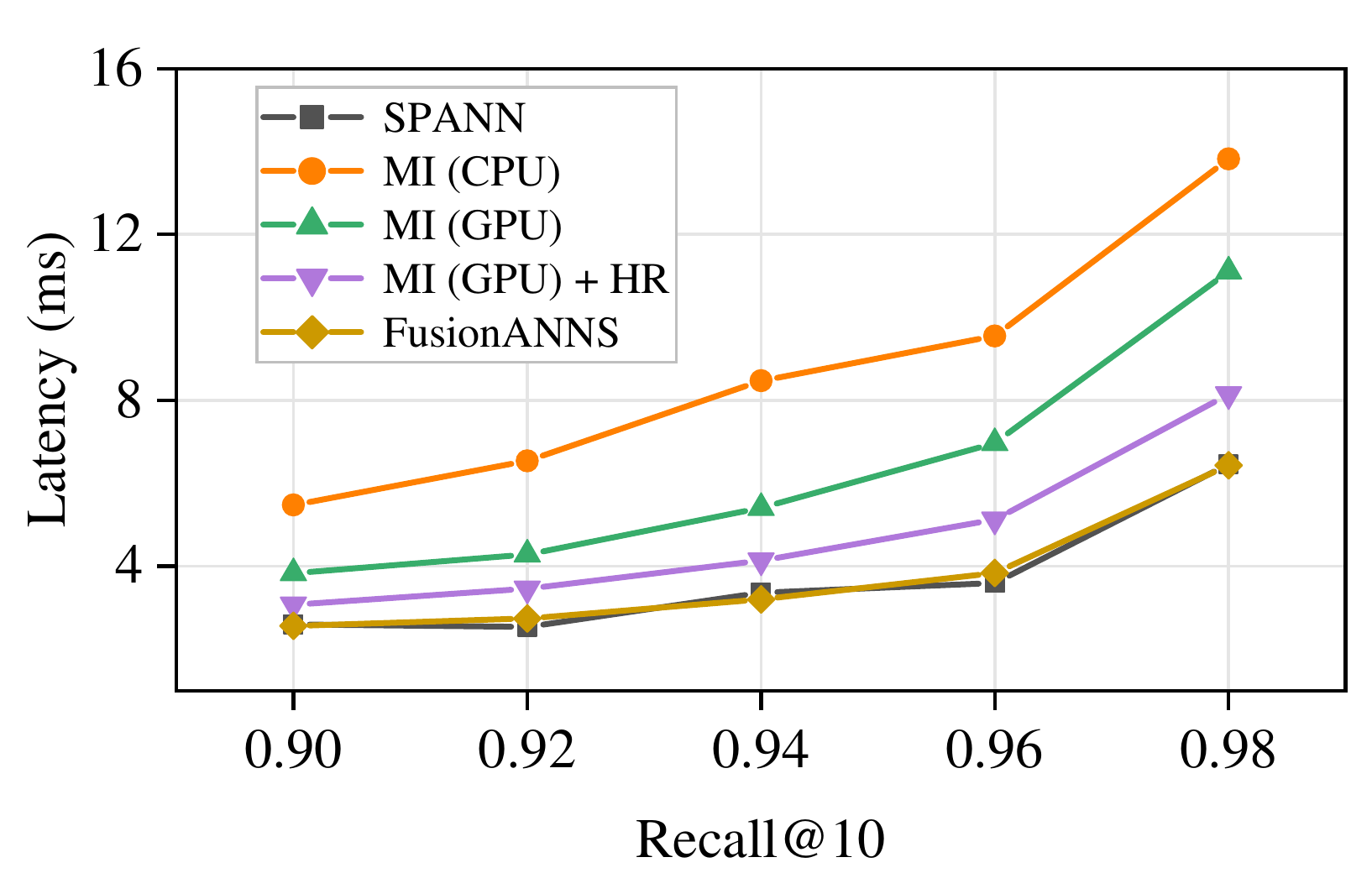}  
        \label{breakdownlatency}  
      } 
      \subfloat[I/O Reduction]{  
        \includegraphics[clip, width=0.32\linewidth]{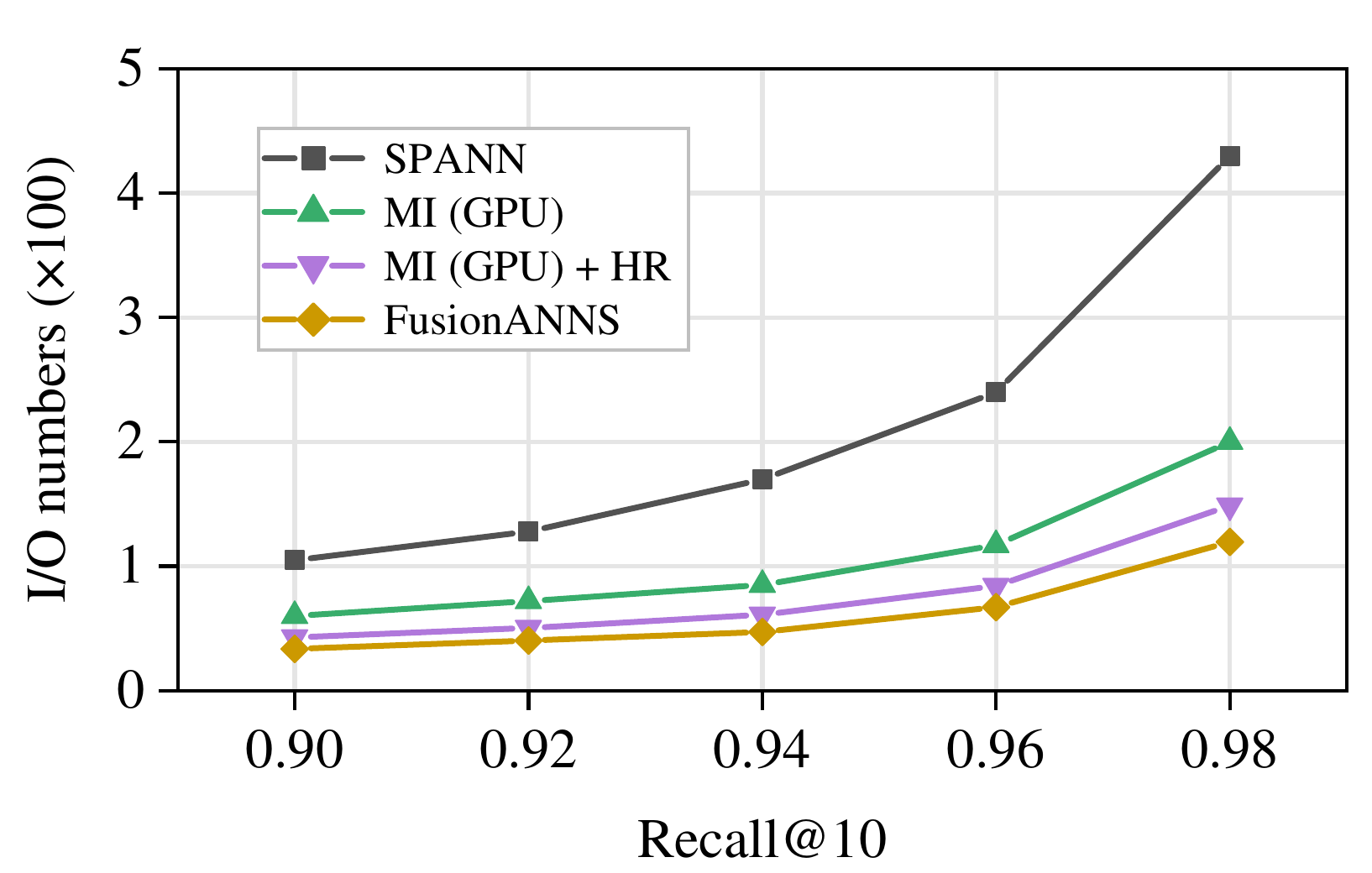}  
        \label{breakdownIOnumbers}  
      }  
      \vspace{-1ex}  
      \caption{Performance improvement and  I/O reduction introduced by different technologies. MI (CPU) and MI (GPU) denote our multi-tiered indexing approach using CPUs and GPU acceleration, respectively. HR denotes heuristic re-ranking. FusionANNS exploits all three techniques, including multi-tiered indexing, heuristic re-ranking, and redundant-aware I/O deduplication.
      } 
      \label{breakdown}
      \vspace{-3ex} 
      \end{figure*} 

\subsection{Effectiveness of Individual Techniques}
In this subsection, we evaluate the effectiveness of individual techniques in FusionANNS using SIFT1B. We first use our multi-tiered indexing technique to conduct a CPU-based variant (i.e., MI(CPU)) in which we replace the GPU with host CPUs to process compressed vectors. Then, we incrementally add other techniques of FusionANNS to evaluate their impacts on the performance and the number of I/Os.

As shown in Figure~\ref{breakdownQPS},  the multi-tiered indexing with CPUs, i.e., MI(CPU), achieves 1.5-4.2~\(\times\) higher QPS compared with SPANN, but suffers from very high latency. However, the multi-tiered indexing with GPU, i.e., MI(GPU), can significantly reduce latency compared with MI(CPU). It also improves the QPS by 5.9-6.8~\(\times\) compared with SPANN. 
This is because CPUs suffer from high latency DRAM accesses, while the high bandwidth and parallel memory access capabilities of GPU's HBM allow MI(GPU) to process PQ distance calculations efficiently. Moreover, multi-tiered indexing enables CPUs only transfer lightweight vector-IDs to the GPU, alleviating the performance bottleneck due to PCIe bandwidth contention.   Based on the multi-tiered indexing, both the heuristic re-ranking (HR) and the redundancy-aware I/O deduplication can further reduce latency and improve QPS by up to 39\% and 17\%, respectively. 

Figure~\ref{breakdownIOnumbers} shows the average I/O numbers aroused by each query. Obviously, the number of I/O requests launched by SPANN increases significantly when a higher accuracy level should be guaranteed. The multi-tiered indexing technique can reduce I/O numbers by 3.2-3.8~\(\times\) compared with SPANN. The heuristic re-ranking and the redundancy-aware I/O deduplication can further reduce I/O numbers by up to 30\% and 23\%, respectively. In addition, each I/O operation launched by SPANN usually involves multiple SSD pages, while other ANNS systems only involve one page. FusionANNS not only significantly reduces the number of I/Os per query, but also reduces the I/O size, thus achieving substantial performance improvement. 




\subsection{Cost and Memory Efficiency}

We compare FusionANNS with other ANNS systems in terms of cost and memory efficiency. We use the QPS/\$ and QPS/GB to evaluate the cost and memory efficiency, respectively. The system cost include the server cost (around \$5000, including CPUs and server chassis), the memory cost (around \$10/GB), the storage cost (\$400 for a 2TB Samsung SSD), and the GPU cost (around \$3000 for Nvidia V100). These prices are referenced from Amazon.  For a fair comparison, we only evaluate FusionANNS, SPANN, and RUMMY because they achieve similar low latency, but exclude DiskANN because it improves QPS at the expense of high latency.

\begin{table}[t]
\centering
\caption{Cost Efficiency (QPS/\$)}
\label{Cost}
\vspace{-2ex} 
\small
\begin{tabular}{@{}cccccc@{}}
\toprule
Datasets & SPANN & RUMMY & \textbf{FusionANNS}   \\ \midrule
SIFT1B   & 0.32   & 0.88   & \textbf{1.98} \\
SPACEV1B & 0.41   & 1.40   & \textbf{2.35} \\
DEEP1B   & 0.12  & 0.15   & \textbf{1.01} \\ \bottomrule
\end{tabular}
\vspace{-1ex}
\end{table}

\begin{table}[t]
\centering
\caption{Memory Efficiency (QPS/GB)}
\vspace{-2ex}
\label{Memory}
\small
\begin{tabular}{@{}cccccc@{}}
\toprule
Datasets & SPANN & RUMMY & \textbf{FusionANNS}   \\ \midrule
SIFT1B   & 29.98   & 47.75   & \textbf{280.23} \\
SPACEV1B & 39.12   & 88.4   & \textbf{330.79} \\
DEEP1B   & 11.17  & 4.51   & \textbf{146.16} \\ \bottomrule
\end{tabular}
\vspace{-3ex}
\end{table}

As shown in Table~\ref{Cost}, FusionANNS achieves 5.67-8.78\(\times\) and 2.25-6.82\(\times\) improvement in QPS/\$ compared to SPANN and RUMMY, respectively. This is mainly due to the significant performance of FusionANNS. Also, FusionANNS achieves higher memory efficiency, as shown in Table~\ref{Memory}. Specifically, for the large-volume dataset DEEP1B, FusionANNS improves memory efficiency by 13.1~\(\times\) and 32.4~\(\times\) compared with SPANN and RUMMY, respectively. FusionANNS dramatically improves the cost and memory efficiency because our multi-level indexing technology can significantly improve performance and reduce memory footprint.


\section{Related Work}

\textbf{In-memory ANNS Solutions.} ANNS has been extensively studied for decades, mainly focusing on in-memory indexing techniques~\cite{IVF_1,IVF_2,HNSW,NSG,SPTAG}. 
\textit{Hierarchical Navigable Small World} (HNSW)~\cite{HNSW} maintains a multi-layered navigable small-world graph in memory to achieve fast ANNS. 
\textit{Space Partition Tree and Graph} (SPTAG)~\cite{SPTAG} exploits a relative neighborhood graph and space partition trees to find several seeds for graph traversal acceleration.
However, existing in-memory ANNS algorithms require a large amount of  memory resource to maintain both raw vectors and their indices. The huge memory requirement significantly increases the
total cost of ownership, impeding the ANNS scaling to
large-scale datasets. FusionANNS leverages large-capacity SSDs to support large-scale datasets, and achieves both high performance and cost efficiency  through multi-tiered indexing-enabled system optimizations.

\textbf{SSD-based ANNS Solutions.} A number of recent studies have proposed hierarchical indices to reduce memory footprint for billion-scale datasets, such as DiskANN~\cite{DiskANN}, SPANN~\cite{SPANN},Starling~\cite{Starling}, BBANN~\cite{BBANN}, and GRIP~\cite{GRIP}. These proposals exploits the characteristics of modern SSDs to optimize vector indexing techniques. SmartANNS~\cite{SmartANNS} explores hierarchical indexing technique for   SmartSSD-based ANNS. It leverages the internal PCIe bandwidth of multiple SmartSSDs to alleviate I/O bottlenecks, achieving near-linear scalability for billion-scale ANNS. However, the throughput of SSD-based ANNS system is quite limited due to severe I/O contention among concurrent queries, making them hard to meet the high-throughput requirement. 
FusionANNS significantly reduces I/Os per query through a novel multi-tiered index structure and CPU/GPU collaborative  searching, and thus achieves both high throughput and low latency.


\textbf{Accelerator-based ANNS Solutions.}
A few recent studies~\cite{CAGRA,GPU_Graph1,GPU_Graph2,GPU_Graph3} exploit GPUs to accelerate the graph traversal involved in graph-based ANNS. 
Moreover, a number of works have exploited PQ techniques~\cite{PQ,IVF-PQ},  GPUs~\cite{JUNO,Faiss_GPU}, and FPGAs~\cite{FPGA_IVFPQ1,FPGA_IVFPQ2,FPGA_IVFPQ3} to accelerate IVF-based ANNS. However, most these approaches can only support small-scale datasets due to the limited memory capacity in GPUs and FPGAs~\cite{CAGRA,GPU_Graph1,GPU_Graph2,GPU_Graph3,FPGA_IVFPQ1}, or can not guarantee high accuracy~\cite{FPGA_IVFPQ1,FPGA_IVFPQ2,FPGA_IVFPQ3}. Although state-of-the-art RUMMY~\cite{RUMMY} expands GPU memory with host memory and proposes a reordered pipelining technique to support billion-scale datasets, its performance is still limited due to extensive data transmission between CPUs and the GPU. 
Through a careful collaboration of  hierarchical indexing, PQ, and GPU acceleration techniques, FusionANNS can eliminate data swapping between GPUs and CPUs, and thus achieves high-throughput and low-latency ANNS for large-scale datasets while still guaranteeing high accuracy.




\vspace{-2ex}

\section{Conclusion}
\vspace{-2ex}
We present FusionANNS, a "CPU + GPU" collaborative processing architecture for billion-scale ANNS. FusionANNS exploits GPU/CPU collaborative filtering and re-ranking mechanisms to significantly improve the performance and  cost efficiency of ANNS while still guaranteeing high accuracy.  Compared with the state-of-the-art SSD-based ANNS solution--SPANN, FusionANNS significantly improves the throughput of concurrent queries while still remaining low latency. Moreover, FusionANNS also achieves higher throughput and cost efficiency than the GPU-accelerated in-memory ANNS solution--RUMMY.

\bibliographystyle{unsrt}
\bibliography{refs}

\end{document}